\documentclass[preprint,aps,prd,floatfix]{revtex4-2}
\usepackage{graphicx}
\usepackage{amsfonts}
\usepackage{amsmath}
\usepackage{rotating}
\usepackage{amssymb,amsthm}

\newtheorem{thm}{Theorem}[section]

\newtheorem{prop}[thm]{Proposition} 
 
\theoremstyle{definition} 
\newtheorem{defn}[thm]{Definition}  
\theoremstyle{remark}

\newcommand{\mbold}[1]{\mbox{\boldmath{\ensuremath{#1}}}}

\def\beq{\begin{eqnarray}}  
\def\eeq{\end{eqnarray}}

\def \bh {\mbox{{\bf h}}}
\def \bomega {\mbox{{\mbold \omega}}}

\begin{document}


\title{A frame based approach to computing symmetries with non-trivial isotropy groups}

\author {D. D. McNutt}
\email{david.d.mcnutt@uit.no}
\affiliation{ Department of Mathematics and Statistics, UiT The Arctic University of Norway, Tromsø, Norway}

\author {A. A. Coley}
\email{alan.coley@dal.ca }
\affiliation{Department of Mathematics and Statistics, Dalhousie University, Halifax, Nova Scotia, Canada, B3H 3J5}

\author{R. J. \surname{van den Hoogen}}
\email{rvandenh@stfx.ca}
\affiliation{Department of Mathematics and Statistics, St. Francis Xavier University, Antigonish, Nova Scotia, Canada, B2G 2W5}



\begin{abstract}

A frame approach to determining the most general solution admitting a desired symmetry  group has been examined previously in Riemannian and teleparallel geometries with some success. In teleparallel geometries, one must determine the general form of the frame and spin connection to generate a general solution admitting the desired symmetry group. Current approaches often rely on the use of the proper frame, where the spin connection is zero. However this leads to particular theoretical and practical problems. In this paper we introduce an entirely general approach to determining the most general Riemann-Cartan geometries which admit a given symmetry group and apply these results to teleparallel geometries. To illustrate the approach we determine the most general geometries, with the minimal number of arbitrary functions, for particular choices of symmetry groups with dimension one, three, six and seven.  In addition, we rigorously show how the teleparallel analogues of the Robertson-Walker, de Sitter and Einstein static spacetimes can be determined.

\end{abstract}

\maketitle

\section{Introduction}

There are a wealth of alternative theories of gravity, including analogues of general relativity (GR), where the torsion tensor and non-metricity tensors are employed in addition to the curvature tensor. In any theory where the torsion tensor or non-metricity tensor are non-zero, the Levi-Civita connection is now a secondary object constructed from the contorsion tensor and the connection which gives rise to the torsion and non-metricity tensors. It is often easier to formulate such theories with their metric tensor and the non Levi-Civita connection as independent quantities. Furthermore, in some theories, such as teleparallel gravity, it is necessary to work with a frame basis instead of a metric tensor. In such theories the role of symmetry is no longer as clearly defined as in pseudo-Riemannian geometry, where the Levi-Civita connection relates solutions to the Killing equations as symmetries, and so we can no longer formulate a symmetry as an isometry of the metric. Instead, for any theory constructed from a metric tensor and a non Levi-Civita connection, the symmetry must satisfy the following conditions

\beq \mathcal{L}_{{\bf X}} g_{ab} = 0 \text{ and } \mathcal{L}_{{\bf X}} \bomega^a_{~b} = 0, \label{Intro1} \eeq

\noindent where the vector field ${\bf X}$ is the infinitesimal generator of the symmetry, $g_{ab}$ is the metric tensor, $\bomega^a_{~b}$ is the connection one-form and $\mathcal{L}$ is the Lie derivative. We note that this definition applies only if the isotropy group of the associated geometry is trivial.

The development of a frame based approach to determining the symmetries of a spacetime has been explored in four-dimensional (4D) GR \cite{chinea1988symmetries, estabrook1996moving, papadopoulos2012locally}. However, the primary obstacle to these approaches is the potential existence of a non-trivial linear isotropy group which is a Lie subgroup of Lorentz frame transformations that leave the associated tensors of the geometry invariant. If a given spacetime has a non-trivial linear isotropy group, determining the group of symmetries requires solving a set of inhomogeneous differential equations with unknown functions, $\lambda^a_{~b}$ \cite{olver1995equivalence}; 

\beq \mathcal{L}_{{\bf X}} \bh^a = \lambda^a_{~b} \bh^b \text{ and } \mathcal{L}_{{\bf X}} \bomega^a_{~b} = \omega^a_{~bc} \lambda^c_{~d} \bh^d, \label{Intro2} \eeq

\noindent where, $\bh^a$ denotes the orthonormal coframe basis and $\lambda^a_{~b}$ is an element of the Lie algebra of Lorentz transformations. We note that the second condition imposes that the vector field ${\bf X}$ is an affine collineation $\mathcal{L}_{{\bf X}} \omega^a_{~bc}=0$. 

 If we have a representation of the symmetry group as a set of vector fields, then solving for the coframe basis and connection is complicated by the unknown functions in $\lambda^a_{~b}$. This problem can be avoided by prolonging to a higher dimensional Lorentzian manifold where the parameters of the basis of the linear isotropy group are now coordinates. While this approach is mathematically well-defined, it is unsatisfactory from a physical perspective where all calculations should be performed on the original manifold.

In the current paper we introduce a new approach to determine the symmetries of any geometry based on a frame bases and an independent connection which admits the torsion tensor and the curvature tensor as geometric objects. In such theories, the connection will be an independent object. For each example, we will also specialize to teleparallel gravity by setting the curvature tensor to zero. We will call any geometry where the non-metricity and curvature tensors vanish a {\it teleparallel geometry}. The approach introduced here relies on the existence of a particular class of invariantly defined frames known as symmetry frames, which is amenable to solving the differential equations arising from \eqref{Intro2}. In this paper, we have assumed the connection is metric compatible in order to discuss Riemann-Cartan geometries. However, the approach outlined in this paper can be readily extended to geometries where the non-metricity tensor is non-zero, i.e., where the connections are no longer metric-compatible.

As an alternative to Riemannian geometries which are typically characterized by the curvature tensor of a Levi-Civita connection calculated from the metric, teleparallel geometries are characterized by the torsion tensor.  Torsion being a function of the coframe, $\bh^a$, derivatives of the coframe, and a zero curvature and metric compatible spin connection. In teleparallel geometries the spacetime indexed representation of the connection, sometimes called the Weitzenb{\"o}ck connection, can be expressed via derivatives of the frame added to the spin connection, \beq \Gamma^{\rho}_{~\nu\mu}=h_c^{~\rho}\left(\partial_\mu h^c_{~\nu}+\omega^c_{~b\mu}h^b_{\nu}\right). \nonumber \eeq
\noindent Assuming an orthonormal frame, the spin connection $\omega^a_{~bc}$, is defined in terms of an arbitrary Lorentz transformation, $\Lambda^a_{~b}$ through the equation,
\beq \omega^a_{~bc} = \Lambda^a_{~d} \bh_c((\Lambda^{-1})^d_{~b}), \label{TP:con} \eeq
\noindent where $\bh_c((\Lambda^{-1})^d_{~b})$ denotes a frame vector acting as a vector field derivation on the components of $(\Lambda^{-1})^d_{~b}$.

Teleparallel geometries provide an alternative framework in which to build a theory of gravity.  A variety of teleparallel gravitational theories based on a Lagrangian can be constructed using various scalars built from the torsion and functions thereof. One subclass of teleparallel gravitational theories is dynamically equivalent to GR and is appropriately called the Teleparallel Equivalent to General Relativity (TEGR) \cite{Aldrovandi_Pereira2013}. A particularly interesting generalization of the TEGR is $F(T)$ teleparallel gravity \cite{Ferraro:2006jd, Ferraro:2008ey, Linder:2010py}. In the {\it covariant} approach to $F(T)$ teleparallel gravity  \cite{Krssak:2018ywd}, the teleparallel geometry is defined in a gauge invariant manner as a geometry with zero curvature, having a spin connection that vanishes in a very special class of frames where all inertial effects are absent, and non-zero in all other frames \cite{Lucas_Obukhov_Pereira2009,Aldrovandi_Pereira2013,Krssak:2018ywd}. Therefore, the resulting teleparallel gravity theory has Lorentz covariant field equations and is therefore locally Lorentz invariant \cite{Krssak_Pereira2015}.

In teleparallel geometries, assuming an orthonormal frame, it is always possible to transform to a frame where the components of the spin connection are zero. Such a frame-connection pair is called a {\it proper frame} and is sometimes desirable when analysing the $F(T)$ field equations. However, a proper frame is not invariantly defined since it is defined in terms of the connection, which is not a tensorial quantity. There have been attempts to use the proper frame to determine the symmetries of a teleparallel geometry,  and for a given group of symmetries determine the largest class of teleparallel geometries which admit this group \cite{pfeifer2022quick, hohmann2019modified, hohmann2021complete}. 

There are two significant problems with using the proper frame to determine symmetries. These problems are related to the fact that the proper frame is not invariantly defined. The first problem is that in the process of solving the equations in \eqref{Intro2}, arbitrary functions of integration will be introduced and it is not possible to determine when such a function is essential to the geometry or if it can be removed by an appropriate coordinate transformation. The second problem is that one cannot determine when two given solutions to \eqref{Intro2} are inequivalent and this can lead to an incomplete list of all geometries which admit a given symmetry group. Our approach differs by choosing a particular class of symmetry frames, that we will define later, which are invariantly defined up to the linear isotropy group and admit a non-zero spin connection. Then by restricting to a particular coordinate system and imposing a chosen vielbein form, all of the remaining frame parameters can be fixed, giving rise to an invariant frame (for this choice of coordinate system). With this class of frames, all  inequivalent geometries admitting a given symmetry group can be determined along with the minimal number of functions in the vielbein and spin connection to specify them. 

We note as well that the proper frame approach cannot predict the existence or non-existence of permitted symmetry groups without extensive calculation. As an example of this, in \cite{Coley:2019zld} it was proven that the only maximally symmetric teleparallel geometry was Minkowski space, and that the (anti-) de Sitter geometries admit, at most, a seven-dimensional symmetry group. This was later verified using the proper frame approach by exhaustively proving that the resulting differential equations from \eqref{Intro2} have no solution \cite{Bahamonde:2021gfp}. In addition, the proper frame approach is applicable only to teleparallel geometries and cannot be used for Riemann-Cartan geometries. 

The outline of the paper is as follows. In the remainder of the introduction we briefly specify the notation used in the paper. In section \ref{sec:FSs} we introduce the class of symmetry frames and define affine frame symmetries for any Riemann-Cartan geometry. In section \ref{sec:FSgroups} we employ symmetry frames to determine the equations required to determine the most general geometries which admit a given symmetry group, and specialize this to teleparallel geometries. In section \ref{sec:Examples}, we provide several examples of this approach by considering the one-dimensional symmetry group, spherical symmetry, Robertson-Walker (RW) symmetry and two seven-dimensional symmetry groups which are special subcases of the RW geometries. In each case, we specialize to teleparallel geometries. In section \ref{sec:Discussion} we summarize our results and discuss further applications. Note that when discussing teleparallel geometries, the associated proper frames will not be discussed.

\subsection{Notation}

We will denote the coordinate indices by $\mu, \nu, \ldots$ and the tangent space indices by $a,b,\ldots$. Unless otherwise indicated the spacetime coordinates will be $x^\mu$. The frame fields are denoted as $\bh_a$ and the dual coframe one-forms are $\bh^a$. The vielbein components are $h_a^{~\mu}$ or $h^a_{~\mu}$. The spacetime metric will be denoted as $g_{\mu \nu}$ while the Minkowski tangent space metric is $\eta_{ab}$. 

To denote a local Lorentz transformation leaving $\eta_{ab}$ unchanged, we write $\Lambda_a^{~b}(x^\mu)$. The spin connection one-form $\bomega^a_{~b}$, is designated by $\bomega^a_{~b} = \omega^a_{~bc} \bh^c$. The curvature and torsion tensors will be denoted, respectively, as $R^a_{~bcd}$ and $T^a_{~bc}$.

Covariant derivatives with respect to a metric-compatible connection will be denoted using a semi-colon, $T_{abc;e}$. When considering the integral curves of a vector field ${\bf X}$, with parameter $\tau$, given as $\phi_{\tau} : M \to M$ we will write the corresponding Lorentz frame transformation as $\phi_\tau^{*} \bh^a = \Lambda^a_{(\tau)~b} \bh^b$.

%


\section{Affine frame symmetries with isotropy} \label{sec:FSs}

The starting point for the determination of symmetries of a Riemann-Cartan geometry, $M$, is the Cartan-Karlhede (CK) algorithm \cite{Coley:2019zld, fonseca1996algebraic}. In this algorithm, a frame is chosen and the parameters of the Lorentz frame transformations are fixed by normalizing the components of the curvature tensor, torsion tensor and their covariant derivatives in an invariant manner. This algorithm provides a wealth of invariants and these can be used to uniquely characterize a geometry locally and determine the dimension of the symmetry group. 

Two useful sequences of discrete invariants are particularly important: the dimension of the linear isotropy group, $dim~H_{p}$ at each iteration, $p$, and the number of functionally independent components of the curvature tensor, torsion tensor and their covariant derivatives up to order $p$, denoted as $t_p$. The linear isotropy group, $H_p$ is a subgroup of the Lorentz frame transformations that leave the curvature tensor, the torsion tensor and their covariant derivatives up to $p^{th}$ order invariant. If $H_p$ is the trivial group, we say that the frame is now an invariant frame. The algorithm ends when $t_{p-1} = t_p$ and $dim~H_{p-1} = dim~H_p$, which indicates that all necessary and sufficient geometric information for the geometry has been produced. 

At the conclusion of the algorithm, say at the $q^{th}$ iteration, the linear isotropy group $H_q$ encodes the freedom in specifying an invariantly defined (co)frame. That is, the CK algorithm produces an invariantly defined (co)frame, $\bh^a$, up to linear isotropy $H_q$ such that for any diffeomorphism, $\phi: M \to M$, the invariantly defined (co)frame satisfies:

\beq \phi^{*} \bh^a = \Lambda^a_{~b} \bh^b \label{pullback}, \eeq

\noindent where $\Lambda^a_{~b} $ belongs to $H_q$ and arises from the coordinate form of $\phi$, i.e., $y^\mu = \phi^\mu(x^\nu)$ \cite{olver1995equivalence}. If $\phi = \phi_\tau$ is generated by exponentiating the vector field locally representing an infinitesimal generator of a symmetry, ${\bf X}$, with some parameter $\tau$, then we can calculate the effect of the Lie derivative of ${\bf X}$ on the frame  as 

\beq \mathcal{L}_{{\bf X}} \bh^a = \lambda^a_{~b} \bh^b, \label{Liederivative:frame} \eeq

\noindent where $\lambda^a_{~b}$ is the Lie algebra generator for $\Lambda^a_{(\tau)~b}$. By using an invariantly defined frame, up to the linear isotropy group, we have reduced the number of unknown functions in $\lambda^a_{~b}$ and restricted their functional dependence to only those coordinates which are affected by $\phi_{\tau}$. To distinguish between those coordinates which are affected or unaffected by the symmetry generators we will denote these as $x^{\alpha'}$ and $x^{\tilde{\alpha}}$, respectively.

This is one part of the recipe for determining or fixing a symmetry in a Riemann-Cartan geometry. Such manifolds are equipped with a metric-compatible connection one-form $\bomega^a_{~b} = \omega^a_{~bc} \bh^c$, where $\bomega_{ab} = - \bomega_{ba}$. If ${\bf X}$ is an infinitesimal generator of a symmetry for the geometry then it must also be an affine collineation \cite{aaman1998riemann}: 
\beq \mathcal{L}_{{\bf X}} \omega^a_{~bc} =0, \label{Liederivative:Con} \eeq

\noindent or equivalently, by using the Leibniz rule for the Lie derivative, this condition can be restated for the connection one-form $\bomega^a_{~b} = \omega^a_{~bc} \bh^c$, 
\beq \mathcal{L}_{{\bf X}} \bomega^a_{~b} = \omega^a_{~bc} \mathcal{L}_{{\bf X}} \bh^c =  \omega^a_{~bc} \lambda^c_{~d} \bh^d. \eeq

\noindent We note that this condition can be rewritten as equation (40) in \cite{hohmann2019modified} using the identity in equation \eqref{SFcon}. From the conditions \eqref{Liederivative:frame} and \eqref{Liederivative:Con} it follows that the Lie derivative of the curvature tensor, the torsion tensor and its covariant derivatives with respect to ${\bf X}$ will vanish. We will call any vector field satisfying equations \eqref{Liederivative:frame} and \eqref{Liederivative:Con} an {\it affine frame symmetry}. We note that if $\lambda^a_{~b} = 0$ then we recover the original definition of an affine frame symmetry in \cite{Coley:2019zld} where the linear isotropy group is trivial and the frame resulting from the CK algorithm is an invariant frame.

While the above definition is necessary and sufficient to determine a symmetry of any Riemann-Cartan geometry, in practice it is difficult to implement as the parameters of the Lorentz transformations must be determined in an invariant manner for each of the possible canonical forms of the curvature tensor and torsion tensor. We will expand this definition to a larger class of frames that are acted on by the Lie algebra generators of the isotropy group under Lie differentiation with the affine frame symmetry generators.

\begin{defn} \label{defn:SymFrame}
Consider a Riemann-Cartan geometry, $({\bh}^a, \bomega^a_{~b})$, admitting an affine frame symmetry, ${\bf X}$. The class of {\bf symmetry frames}, $\bh^a$, satisfy the following condition

\beq && \mathcal{L}_{{\bf X}} \bh^a = f_{X}^{~\hat{i}} \lambda^a_{\hat{i}~b} \bh^b, \label{TP:frm:sym} \eeq

\noindent where $\lambda^a_{\hat{i}~b}$ are basis elements of the Lie algebra of the isotropy group (so that $\lambda_{\hat{i} ab} =- \lambda_{\hat{i} ba}$) and $\hat{i}$ ranges from 1 to the dimension of the isotropy group. The components of $f_X^{~\hat{i}}$ are functions dependent on the coordinates affected by the symmetry generators $x^{\alpha'}$ and have been introduced to permit the possibility of an arbitrary infinitesimal generator of an element of the isotropy group. This is a general definition and is not (strictly speaking) an invariantly defined frame until the components $f_{ X}^{~\hat{i}}$ are fixed in some coordinate independent way.

\end{defn}

Assuming that equations \eqref{Liederivative:Con} and \eqref{TP:frm:sym} are satisfied, so that ${\bf X}$ generates an affine frame symmetry, we find that for $g_{\mu \nu} = \eta_{ab} \bh^a_{~\mu} \bh^b_{~\nu}$, 

\beq \mathcal{L}_{{\bf X}} (\eta_{ab} \bh^a \bh^b) = f_X^{~\hat{i}} [ \lambda_{\hat{i} bc} \bh^b \bh^c + \lambda_{\hat{i} cb} \bh^b \bh^c ] = 0, \eeq

\noindent which ensures that ${\bf X}$ is a Killing vector field as well.

The class of symmetry frames is guaranteed to exist since they are constructed by taking the class of frames generated from the CK algorithm, which are invariantly defined up to the linear isotropy group, and acting upon them by the group of Lorentz transformations that leave the representation of the isotropy group unchanged (which we have denoted as $\overline{Iso}$). This can be determined using the chosen Lie algebra basis for the Lorentz group and picking those Lie algebra generators whose commutator with the Lie algebra elements of the isotropy group lie in the isotropy group. To be precise, we can determine the Lie algebra generators of $\overline{Iso}$, $\tilde{\lambda}$, by solving the $n$ equations, $[\lambda_{\hat{i}}, \tilde{\lambda}] \in~ \text{span}\{ \lambda_{\hat{1}}, \ldots, \lambda_{\hat{n}}\}$, and then exponentiating the linearly independent parts of $\tilde{\lambda}$ to produce $\overline{Iso}$. For example, in the case of spherical symmetry, there are three generators of the isotropy group, ${\bf X}_I$, $I=1,2,3$ and the group $\overline{Iso}$ consists of all spatial rotations.

Since the components of $f_X^{~\hat{i}}$ are associated with the Lie derivative of a symmetry generating vector field, they are tensor quantities that depend on the frame. Under a change of frame, $\tilde{\bh}^a = \tilde{\Lambda}^a_{~b} \bh^b$, the components transform as: 
\beq {\bf X}_I ( \tilde{\Lambda}^a_{~b}) [\tilde{\Lambda}^{-1}]^b_{~c} + \tilde{\Lambda}^a_{~b} f_I^{~\hat{i}} \lambda^b_{\hat{i}~d} [\tilde{\Lambda}^{-1}]^d_{~c} = \tilde{f}_I^{~\hat{i}} \lambda^a_{\hat{i}~ c}. \label{frot} \eeq
\noindent We can employ them to construct invariantly defined frames up to some minimal subgroup of $\overline{Iso}$ or invariant frames if this subgroup is trivial. To be precise, if we employ the frame freedom in $\overline{Iso}$ to fix the form of the $f_{ X}^{~\hat{i}}$ as much as possible, so that some elements are constant or zero, the symmetry frame is then an {\it invariantly defined symmetry frame up to the isotropy group $\overline{H}_q$}, where $\overline{H}_q$ is some subgroup of $\overline{Iso}$ which leaves the chosen form of $f^{~\hat{i}}_{X}$ unchanged. Again, using the spherically symmetric case as an example, a choice of $f^{~\hat{i}}_{{X}_I} \equiv f^{~\hat{i}}_I$, $I=1,2,3$, for each symmetry generator can be made so that this subgroup, $\overline{H}_q$, consists of rotations in the $\bh^3-\bh^4$ plane where their Lorentz parameters are constant. In subsection \ref{sec:Examples}, we will outline the procedure to fix the components of $f_X^{~\hat{i}}$ and determine $\overline{H}_q$ for the axially-symmetric, spherically symmetric and spatially homogeneous and isotropic symmetric geometries, respectively.

We note that once a class of symmetry frames is invariantly defined up to the group $\overline{H}_q$, any additional fixing of the remaining frame parameters should be maximally fixed, up to the linear isotropy group of the CK algorithm, $H_q$, in an invariant manner. This can be done using the curvature tensor, the torsion tensor or their respective covariant derivatives, as in the CK algorithm. While this can be done readily for a given geometry, or even particular subclasses of geometries, it is difficult to implement for a general class of geometries. However, since the solution to equation \eqref{TP:frm:sym} yields a particular frame ansatz, we can sidestep the problem of fixing the frame invariantly by restricting the set of permitted diffeomorphisms of the geometry to the ``shape preserving" transformations, those which preserve the function form of the metric or coframe \cite{swift1986everywhere} but alter the expressions for the metric or vielbein functions. That is, we can make a choice of vielbein, which provides a way to further fix the frame fully and construct an invariant frame, relative to the given coordinate system. This is sufficient for finding inequivalent solutions but leads to difficulties if they are to be subclassified.

\section{Determining the most general geometry for a given isometry group} \label{sec:FSgroups}

Let us assume that we have a Lie group of affine frame symmetries of dimension $N$, with a non-trivial isotropy group with dimension $n$ $(n < N)$. We choose coordinates so that the symmetry group is represented as a set of vector fields, ${\bf X}_I$, where $I, J, K \in [1, N]$, with the corresponding Lie algebra: 

\beq [{\bf X}_I, {\bf X}_J] = C^K_{~IJ} {\bf X}_K, \label{CijkSym} \eeq

\noindent where $C^K_{~IJ}$ are structure constants of the Lie algebra.

Then we can determine the most general Riemann-Cartan geometries admitting this Lie group as affine symmetries by solving the following equations for an orthonormal symmetry coframe \footnote{A null, or even complex null frame will work as well, depending on the desired representation of the isotropy group}, $\bh^a$,  the associated metric, $g_{\mu \nu} = \eta_{ab} h^a_{~\mu} h^b_{~\nu}$, and with a metric-compatible connection, $\omega^a_{~bc}$: 
\beq &&\mathcal{L}_{{\bf X}_I} \bh^a = f_I^{~\hat{i}} \lambda^a_{~\hat{i}~b} \bh^b \label{TP:sym1} \\ 
&& \mathcal{L}_{{\bf X}_I} \omega^a_{~bc} = 0 , \label{TP:sym2} \eeq

\noindent where $\hat{i}, \hat{j}, \hat{k} \in [1,n]$ and the components of $f_I^{~\hat{i}}$ are functions of the coordinates, $x^{\alpha'}$. The form of $f_I^{~\hat{i}}$ can be specified by choosing some components to be equal to others, to be constant or to be zero. This defines a class of symmetry frames invariantly up to $\overline{H}_q$. For the moment we will consider an arbitrary symmetry frame. In addition, due to the properties of the Lie derivative, we have an additional relationship coming from \eqref{CijkSym}:
\beq [\mathcal{L}_{{\bf X}_I}, \mathcal{L}_{{\bf X}_J}] \bh^a = \mathcal{L}_{[{\bf X}_I, {\bf X}_J] } \bh^a = C^K_{~IJ} \mathcal{L}_{{\bf X}_K} \bh^a. \label{TP:sym3} \eeq

We will focus for the moment on equations \eqref{TP:sym1} and \eqref{TP:sym3}. These can be expanded to give the following equations:
\beq X_I^{~\nu} \partial_{\nu} h^a_{~\mu} + \partial_{\mu} {X}_I^{~\nu} h^a_{~\nu} = f_I^{~\hat{i}} \lambda^a_{\hat{i}~b} h^b_{~\mu}, \label{TP:sym1big} \eeq

\noindent and 
\beq 2 {\bf X}_{[I} ( f_{J]}^{~\hat{k}}) - f_I^{~\hat{i}} f_J^{~\hat{j}} C^{\hat{k}}_{~\hat{i} \hat{j}} = C^K_{~IJ} f_K^{~\hat{k}}, \label{TP:sym3big} \eeq

\noindent where the associated Lie algebra of the isotropy group is given as: \beq [\lambda_{\hat{i}}, \lambda_{\hat{j}}] = \lambda^a_{\hat{i}~b} \lambda^b_{\hat{j}~c} - \lambda^a_{\hat{j}~b} \lambda^b_{\hat{i}~c} =  C^{\hat{k}}_{~\hat{i} \hat{j}} \lambda^a_{\hat{k}~c}. \nonumber \eeq

These equations determine the {\bf most general} frame basis which admits a given symmetry group with a non-trivial isotropy group. To determine the connection we must solve equation \eqref{TP:sym2} relative to this frame. Using the coordinate basis expression for the Lie derivative of the connection in  \cite{yano2020theory}, the corresponding frame basis expression is,

\beq \mathcal{L}_{{\bf X}_I}(\omega^a_{~bc}) \bh_a = {\bf R}( {\bf X},\bh_c) \bh_b - {\bf T}({\bf X}, \nabla_c \bh_b) + \nabla_c {\bf T}({\bf X}, \bh_b) + \nabla_c \nabla_b {\bf X} - \nabla_{\nabla_c \bh_b} {\bf X}, \eeq

\noindent where ${\bf R}(\bh_d, \bh_c)\bh_b = R^a_{~bcd} \bh_a$ and ${\bf T} ( \bh_c, \bh_b) = T^a_{~bc} \bh_a$, respectively, denote the curvature tensor and torsion tensor of the connection $\omega^a_{~bc}$, and $\nabla_a = \nabla_{\bh_a}$. 

We can use equations \eqref{TP:sym1} and \eqref{TP:sym2} to find the following set of equations for $\omega^a_{~bc}$:

\beq \mathcal{L}_{{\bf X}_I}(\omega^a_{~bc}) =  { X}_I^{~d} \bh_d( \omega^a_{~bc}) + \omega^d_{~bc} f_I^{~\hat{i}} \lambda^a_{\hat{i}~d} - \omega^a_{~dc} f_I^{~\hat{i}} \lambda^d_{\hat{i}~b} - \omega^a_{~bd} f_I^{~\hat{i}} \lambda^d_{\hat{i}~c} - \bh_c( f_I^{~\hat{i}}) \lambda^a_{\hat{i}~b} = 0 \label{SFcon} \eeq

\noindent where $\bh_c(f(x^\alpha))$ denotes the frame derivative of a function $f(x^\alpha)$ with respect to $\bh_c$. 

\begin{thm} 
The most general Riemann-Cartan geometry admitting a given group of symmetries, ${\bf X}_I,~ I,J,K \in \{1, \ldots, N\}$ with a non-trivial isotropy group of dimension $n$ can be determined by solving for the unknowns $h^a_{~\mu}$, $f_I^{~\hat{i}}$ (with $\hat{i}, \hat{j}, \hat{k} \in \{1,\ldots, n\}$) and $\omega^a_{~bc}$ from the following equations:

\beq \begin{aligned} 
& { X}_I^{~\nu} \partial_{\nu} h^a_{~\mu} + \partial_{\mu} { X}_I^{~\nu} h^a_{~\nu} = f_I^{~\hat{i}} \lambda^a_{\hat{i}~b} h^b_{~\mu} \\
& 2{\bf X}_{[I} ( f_{J]}^{~\hat{k}}) - f_I^{~\hat{i}} f_J^{~\hat{j}} C^{\hat{k}}_{~\hat{i} \hat{j}} = C^K_{~IJ} f_K^{~\hat{k}} \\
& { X}_I^{~d} \bh_d( \omega^a_{~bc}) + \omega^d_{~bc} f_I^{~\hat{i}} \lambda^a_{\hat{i}~d} - \omega^a_{~dc} f_I^{~\hat{i}} \lambda^d_{\hat{i}~b} - \omega^a_{~bd} f_I^{~\hat{i}} \lambda^d_{\hat{i}~c} - \bh_c( f_I^{~\hat{i}}) \lambda^a_{\hat{i}~b} = 0. \end{aligned} \label{Sym:RC:Prop} \eeq

\noindent where $\{ \lambda^a_{\hat{i}~b}\}_{\hat{i}=1}^n$ is a basis of the Lie algebra of the  isotropy group, $[\lambda_{\hat{i}}, \lambda_{\hat{j}}] = C^{\hat{k}}_{~\hat{i}\hat{j}} \lambda_{\hat{k}}$, $[{\bf X}_I, {\bf X}_J] = C^K_{~IJ} {\bf X}_K$. 

\end{thm}

\noindent We remark that while the above equations apply to any symmetry frame, solving these equations can be more easily attained by using an invariantly defined symmetry frame up to $\overline{H}_q$ where the components of $f_I^{~\hat{i}}$ have been invariantly specified. As discussed in the previous subsection, this entails using the group $\overline{Iso}$ to fix the form of the matrix $f_I^{~\hat{i}}$ in an invariant manner, for example by setting some components to be constant, to be zero or equal to other components.

To specialize this result to a teleparallel geometry, in addition to the differential equation coming from equation \eqref{Liederivative:Con}, the connection must be flat and so the  curvature tensor expressed in terms of the spin connection and its derivatives must vanish. This condition is equivalent to \eqref{TP:con} for some choice of Lorentz transformation $\Lambda^a_{~b}$. In summary:

\begin{thm} 
The most general teleparallel geometry which admits a given group of symmetries, ${\bf X}_I,~ I,J,K \in \{1, \ldots N\}$ with a non-trivial isotropy group of dimension $n$ can be determined by solving for the unknowns $h^a_{~\mu}$, $f_I^{~\hat{i}}$ (with $\hat{i}, \hat{j}, \hat{k} \in \{1, \ldots n\}$) and $\omega^a_{~bc}$ from equations \eqref{Sym:RC:Prop} along with an additional condition:

\beq \begin{aligned} 
& R^a_{\phantom{a}bcd} = h_c^{\mu} \partial_\mu \omega^a_{\phantom{a}bd}- h_d^{\nu}\partial_\nu \omega^a_{\phantom{a}bc}+\omega^a_{\phantom{a}fc}\omega^f_{\phantom{a}bd}-\omega^a_{\phantom{a}fd}\omega^f_{\phantom{a}bc} = 0. \end{aligned} \label{Sym:Prop} \eeq

\end{thm}

\section{Examples} \label{sec:Examples}

\subsection{One-dimensional symmetry groups}

If the symmetry group is one-dimensional, one may always choose a neighbourhood and rectify the infinitesimal generator, ${\bf X}$, so that it is of the form ${\bf X} = \partial_\chi$ with coordinates $x^\mu = (x^{\tilde{\alpha}}, \chi)$ with $\tilde{\alpha} = 1,2,3$. We note that the coordinate $\chi$ can be spacelike, timelike or null, depending on the nature of the affine frame symmetry. To distinguish the behaviour of the resulting affine-frame symmetry, we must examine the norm of the vector field ${\bf X}$ and its effect on the symmetry frame:
\beq \mathcal{L}_{{\bf X}} \bh^a = f(\chi) \lambda^a_{~b} \bh^b. \label{1DSF} \eeq

\noindent where $f(\chi)$ is an arbitrary function to be determined. If $f$ is identically zero, the resulting affine frame symmetry is translational, while if $f$ is non-trivial, the affine frame symmetry is an affine frame isotropy. By computing the squared norm of $\lambda^a_{~b}$ we can use global Lorentz transformations to simplify the form of this anti-symmetric matrix into four classes: a boost, a null rotation, a loxodromic transformation or a spatial rotation. Following \cite{hall2004symmetries} we will use two-forms to describe the matrices $\lambda_{ab} = \eta_{ac}\lambda^c_{~b}$ so that these transformations are described relative to an orthonormal frame, respectively, as:

\beq \lambda_{ab} \in \{ \bh^1 \wedge \bh^4,~\bh^1 \wedge \bh^2 + \bh^2 \wedge \bh^4, \bh^1 \wedge \bh^4 + a_0 \bh^2 \wedge \bh^4, \bh^2 \wedge \bh^3 \}. \label{1DLieAlg} \eeq

\noindent where $a_0$ is a real-valued constant acting as a fixed parameter in the loxodromic transformation. In this general setting, determining $\overline{Iso}$ and $\overline{H}_q$ for each of the possible Lie algebra generators in this representation is a tedious but straightforward calculation. We will employ axisymmetry in order to illustrate this approach in section \ref{sec:1dexample} .

Without loss of generality, we can use a Lorentz transformation to eliminate the arbitrary function $f$ using $\tilde{\Lambda}^a_{~b} = [e^{\tau {\bf \lambda}}]^a_{~b}$ and requiring that the parameter $\tau$ is a function of $\chi$ determined by equation \eqref{frot}. Explicitly, we will solve the following differential equation:

\beq X(\tilde{\Lambda}^a_{~b}) [\tilde{\Lambda}^{-1}]^b_{~c} + \tilde{\Lambda}^a_{~b} f \lambda^b_{~e} [\tilde{\Lambda}^{-1}]^e_{~c} = 0.  \eeq

\noindent Since $\tilde{\Lambda}^a_{~b}$ has the same Lie algebra generator as $\lambda^a_{~b}$, $\lambda^a_{~b}$  in the second term is invariant under this transformation, and the matrix equation reduces to a simple differential equation:

\beq \tau_{,\chi} + f = 0. \nonumber \eeq

\noindent Thus by integrating $\tau$ we can always remove $f$ from the Lie derivative of $\bh^a$ and instead involve it in the connection, according to the transformation rule for the connection one-form: 
\beq \tilde{\bomega}^a_{~b} = \Lambda^a_{~c} \bomega^c_{~d} (\Lambda^{-1})^d_{~b} + \Lambda^a_{~c} (d \Lambda^{-1})^c_{~b}. \eeq

The new equation for the symmetry frame is now 
\beq \mathcal{L}_{{\bf X}} \tilde{\bh}^a = 0, \eeq

\noindent which implies that the vielbein matrix components are independent of $\chi$, i.e., $\bh^a = h^a_{~\alpha}(x^{\tilde{\alpha}}) d x^{\alpha}$. This frame is another symmetry frame, however, it is now invariantly defined up to $\overline{H}_q$ since $f_I^{~\hat{i}}=0$. Imposing the affine frame symmetry conditions on the new connection gives an additional condition on the function $f$. 

In this coordinate system, with the appropriate symmetry frame determined  we can impose the second condition for an affine frame symmetry: 
\beq  \mathcal{L}_{{\bf X}} \omega^a_{~bc} = {\bf X} ( \omega^a_{~bc}) = 0. \eeq
\noindent This is satisfied if and only if the connection coefficients, $\omega^a_{~bc}$, are independent of $\chi$ and so any choice of $\omega^a_{~bc} = W^a_{~bc}(x^{\tilde	{\alpha}})$ is sufficient to define the connection. Using the transformation rule for the connection one-form with the Lorentz transformation $\tilde{\Lambda}^a_{~b}$ with parameter $\tau$ such that $\partial_{\chi} \tau = -f$, Lie differentiation with respect to $\chi$ of this new connection one-form must be zero, and this implies that $\partial_\chi^2 \tau = 0$. Since $\partial_\chi  \tau = -f$, this function is independent of $\chi$, and hence is a constant, $\epsilon$. With the function $f(\chi)$ specified, we have the following result. 

\begin{prop}
The most general Riemann-Cartan geometry admitting a one-dimensional symmetry, ${\bf X} = \partial_{\chi}$ is given by a coframe and a connection of the form 

\beq \bh^a = h^a_{~\alpha}(x^{\tilde{\alpha}}) dx^{\alpha}, \bomega^a_{~b} = W^a_{~bc}(x^{\tilde{\alpha}}) \bh^c - \bh_c( \epsilon \chi) \lambda^a_{~b} \bh^c \label{propiv2eqn} \eeq

\noindent where $\bh_c(\epsilon \chi)$ denotes the derivation of the linear function $\epsilon \chi$ by the frame element $\bh_c$ and $\epsilon = 0, 1$.

The magnitude of the infinitesimal generator $|{\bf X}|$ determines whether the affine frame symmetry is timelike, null or spacelike at each point, while the Lie algebra element $\lambda^a_{~b}$ indicates whether the symmetry is translational, or it admits a fixed point and hence is an isotropy.

\end{prop}

We remark here that, that the second term in the connection one-form in equation \eqref{propiv2eqn} can always be absorbed into the first term. However, where our intention is to construct the most general class of Riemann-Cartan geometries which admit one affine frame symmetry, we have left this second term explicit, in order to emphasise the role of $\lambda^a_{~b}$ in choosing the affine frame symmetry.

We also note that the frame in equation \eqref{propiv2eqn} is invariantly defined (up to $\overline{H}_q$), as this example is very general, we cannot explicitly fix the frame parameters further but point out that it is always possible to construct an invariantly defined frame when the symmetry group is one dimensional.

In the case of teleparallel geometries, there exists a local Lorentz transformation, $\hat{\Lambda}^a_{~b}$, that yields the spin connection through the equation
\beq \bomega^a_{~b} = \hat{\Lambda}^a_{~c} (d \hat{\Lambda}^{-1})^c_{~b}. \eeq
\noindent In this case, it is easiest to formulate the most general frame and connection admitting a single affine frame symmetry by switching to a proper coframe first, i.e., 
\beq {\bf \hat{h}}^a = (\hat{\Lambda}^{-1})^a_{~b} \bh^b, ~\bomega^a_{~b} = {\bf 0} \nonumber \eeq

\noindent Relative to the proper frame, the symmetry generator acts as 
\beq \mathcal{L}_{{\bf X}} {\bf \hat{h}}^a = \lambda^a_{~b} {\bf \hat{h}}^a \nonumber \eeq

\noindent where $\lambda^a_{~b}$ is some arbitrary element of the Lie algebra of the Lorentz group. 

In previous papers it has been shown that $\lambda^a_{~b}$ must have constant components \cite{pfeifer2022quick}. Due to this, it is tempting to use the proper frame as the preferred frame to determine symmetries. However, the determination of proper frames is not invariantly defined using tensor quantities, and so it is difficult to determine when an arbitrary function in a given solution is essential, or only a coordinate artefact. Whereas in our approach, the arbitrary functions in the vielbein describe the minimal number of functions required to be a teleparallel geometry admitting one affine frame symmetry.

If $\lambda^a_{~b}$ vanishes, then the symmetry generator is translational. Otherwise, this constant matrix can be specified using global Lorentz transformations depending on the squared norm of $\lambda^a_{~b}$. If it is negative, positive or zero, the matrix $\lambda^a_{~b}$ can be chosen as a simple generator of a boost, rotation or null rotation, respectively. 

Once this invariant is determined, the above process can be repeated for a teleparallel geometry. Using the fact that we can always pick the proper frame in equation \eqref{1DSF}, we can employ global Lorentz transformations to bring $\lambda^a_{~b}$ into one of the canonical representations for a boost, rotation or a null rotation. Then by applying the Lorentz transformation associated with $\lambda^a_{~b}$ and solving a simple differential equation, it is possible to shift this term to the spin connection so that the Lie derivative of the frame basis with respect to ${\bf X}$ is zero. The role of $\lambda^a_{~b}$ is even more apparent in the coordinate basis as only the $d\chi$ term appears in the connection one-form:

\beq - \bh_c( \epsilon \chi) \lambda^a_{~b} \bh^c = - \epsilon \lambda^a_{~b} d\chi. \label{1sym:con} \eeq

Summarizing these observations, we have the following result.

\begin{prop} \label{prop:IV2}
The most general teleparallel geometry admitting a one-dimensional symmetry, ${\bf X} = \partial_{\chi}$ is given by a frame and a connection of the form 
\beq {\bh}^a = {h}^a_{~\alpha}(x^{\tilde{\alpha}}) d x^{\alpha}, \bomega^a_{~b} =  - \epsilon \lambda^a_{~b} d\chi. \eeq
\noindent where $\bh_c(\epsilon \chi)$ denotes the derivation of the linear function $\epsilon \chi$ by the frame element $\bh_c$ and $\epsilon = 0, 1$.

The magnitude of the infinitesimal generator $|{\bf X}|$ determines whether the affine frame symmetry is timelike, null or spacelike at each point, while the Lie algebra element $\lambda^a_{~b}$ indicates whether the symmetry is translational, or it admits a fixed point and hence is an isotropy.

\end{prop}

\noindent As in the Riemann-Cartan case, the frame is invariantly defined up to $\overline{H}_q$. In each example of a one dimensional symmetry group, an invariantly defined symmetry frame can be found. 

The frame and spin connection given here is the most general as the majority of the spin connection has been absorbed into the frame which consists of fully arbitrary components dependent on $x^{\tilde{\alpha}}$. The remaining non-zero part of the spin connection can only be removed using a Lorentz transformation with $\chi$-dependent parameters and the resulting symmetry frame is no longer invariantly defined. 

\subsubsection{An application: Axisymmetric geometries} \label{sec:1dexample}

As an example that will be relevant in the remaining examples, we will consider the class of geometries which admit rotational isotropy. These have been discussed previously for teleparallel solutions, either by using the axially symmetric field equations \cite{nashed2010stationary} or by using the proper frame approach \cite{hohmann2019modified}. 

In \cite{hohmann2019modified}, two solutions were found to admit axial symmetry. One choice relied on a non-trivial homomorphism from $SO(2)$ into $SO(1,3)$, while the second arose from the trivial homomorphism from $SO(2)$ into $SO(1,3)$. The solution corresponding to the trivial homomorphism is not axisymmetric, in the sense of admitting a symmetry in the isotropy group, but instead admits a translational affine frame symmetry, which can be seen by noting that the proper frame is already a symmetry frame with $\lambda^a_{~b} = 0$. 

Our approach recovers both cases as solutions where, relative to the appropriate symmetry frame, the connection one-form in equation \eqref{1sym:con} is either non-trivial or identically zero, respectively. This provides an explanation for why the second class of frames corresponding to the trivial homomorphism in \cite{hohmann2019modified} does not have a proper limit to spherically symmetric geometries. In order for this to occur, the affine-symmetry generator $\partial_{\phi}$ must lie in the isotropy group in the context of spherical symmetry but the vanishing of the connection one-form, relative to the invariantly defined symmetry frame, implies that $\partial_{\phi}$ does not belong to the isotropy subgroup and hence cannot be expanded into the spherically symmetric group.

For these axisymmetric geometries we will work in the coordinate system $x^\mu = (t,r,\theta,\phi)$ and treat the vector field ${\bf X} = \partial_{\phi}$ as the generator of an axisymmetric symmetry. The Lie algebra generator is then 

\beq \lambda^a_{~b} = \left[ \begin{array}{cccc} 0 & 0 & 0 & 0 \\ 0 & 0 & 0 & 0 \\ 0 & 0 & 0 & 1 \\ 0 & 0 & -1 & 0 \end{array} \right] \label{axilambda} \eeq

\noindent which corresponds to the representation $\lambda = \bh^3 \wedge \bh^4$. 

With this choice, we can determine the frame transformations that leave this form invariant by solving for $\tilde{\lambda}$ from $[\lambda, \tilde{\lambda}]= \lambda$ and exponentiating $\tilde{\lambda}$ to determine the elements of the Lorentz frame transformation group.   $\overline{Iso}$ consists of rotations in the $\bh^3-\bh^4$ plane, and boosts in the $\bh^1-\bh^2$ subspace. As in the general case of a Riemann-Cartan geometry, the function $f(\phi)$ can be set to zero, which fixes the frame invariantly. To determine $\overline{H}_q$, we solve equation \eqref{frot} with $f^{\hat{i}}_I= \tilde{f}^{\hat{i}}_I = 0$ for $\tilde{\Lambda}^a_{~b}$ in $\overline{Iso}$. This implies that \beq {\bf X}_I(\tilde{\Lambda}^a_{~b}) = 0 = \partial_{\phi}(\tilde{\Lambda}^a_{~b}) = 0, \nonumber \eeq
\noindent and so $\overline{H}_q$ consists of elements of $\overline{Iso}$ but with constant parameters.

The remaining frame freedom can be fixed by either using the curvature and torsion tensor or by employing shape-preserving transformations to simplify the vielbein. Due to the large degree of generality in the vielbein, we will omit this step, as further frame fixing will not reduce the number of arbitrary functions. However, we note that it is always possible to construct an invariantly defined frame using either approach for a given solution.

Relative to a generic coframe, where the vielbein are arbitrary but independent of $\phi$:

\beq \bh^a = h^a_{~\beta}( x^{\tilde{\alpha}}) dx^{\beta}, \eeq   

\noindent with $x^{\tilde{\alpha}} = (t,r,\theta)$. To ensure the traditional axisymmetric geometry, the norm of ${\bf X}$ must be spatial and this gives an inequality for the vielbein, 

\beq g_{\alpha \beta} X^{\alpha} X^{\beta} = \eta_{ab} h^a_{~\alpha} h^b_{\beta} X^{\alpha} X^{\beta} = h^a_{~\phi} h_{a\phi} > 0, \label{axi:norm} \eeq

\noindent in a neighbourhood of the axis, and zero only at points on the axis. The connection one-form for a Riemann-Cartan geometry then takes the form

\beq \bomega^a_{~b} = W^a_{~bc}( x^{\tilde{\alpha}}) \bh^c - \bh_c (\epsilon \phi) \lambda^a_{~b} \bh^c, \eeq

\noindent where $\epsilon = 0, 1$

In the case of a teleparallel geometry, we can employ the same argument leading to Proposition \ref{prop:IV2} to describe an axisymmetric teleparallel geometry in terms of the following coframe and connection one-form: 

\beq \bh^a = h^a_{~\beta}( x^{\tilde{\alpha}}) dx^{\beta} \text{ and } \bomega^a_{~b} = - \bh_c ( \epsilon \phi) \lambda^a_{~b} \bh^c = - \epsilon \lambda^a_{~b} d\phi, \eeq

\noindent where $\lambda^a_{~b}$ is given in equation \eqref{axilambda} and the vielbein must satisfy equation \eqref{axi:norm}.

Within the class of axisymmetric geometries defined in Riemann-Cartan geometries with non-zero torsion, the stationary axisymmetric solutions are particularly important from a physical perspective as these will contain solutions similar to the Kerr metric. To the authors' knowledge these geometries have not been investigated using symmetry methods \cite{nashed2010stationary}. Relative to this coordinate system where the symmetry generator takes the form, ${\bf X} = \partial_{\phi}$, stationary axisymmetric geometries can be generated by the inclusion of a second affine frame symmetry, ${\bf Y}$ with $Y^a Y_a < 0$ locally, which commutes with ${\bf X}$; i.e., $[{\bf X}, {\bf Y}] =0$. A new coordinate system can be chosen so that ${\bf X} = \partial_{\phi}$ and ${\bf Y} = \partial_{t}$. The timelike condition on ${\bf Y}$ imposes an additional condition on the vielbein. 

In a similar manner to the stationary axisymmetric geometries, a rigorous investigation of the static axisymmetric geometries in Riemann-Cartan geometries with non-zero torsion has not been performed. However, within the current framework, it is not difficult to formulate these geometries. Static axisymmetric geometries arise by requiring that ${\bf Y}$ is hypersurface orthogonal and hence satisfies 
\beq {\bf Y} \wedge d {\bf Y} = 0. \eeq
The larger symmetry group introduces an additional function, $f_I^{~\hat{i}}$, into equation \eqref{TP:sym1} and this can be determined using equation \eqref{TP:sym3}. The analysis of these equations is not difficult and lead to the condition that the vielbein and the components of the connection one-form must be independent of the coordinates $t$ and $\phi$.
 
\subsection{A three-dimensional symmetry group: spherical symmetry} 

Working in coordinates $ x^\mu = (t, r, \theta, \phi)$, the affine frame symmetry generators associated with spherical symmetry are,
\beq \begin{aligned} & {\bf X}_z = \partial_{\phi},~{\bf X}_y = - \cos \phi \partial_{\theta} + \frac{\sin \phi}{\tan \theta} \partial_{\phi}, {\bf X}_x = \sin \phi \partial_{\theta} + \frac{\cos \phi}{\tan \theta} \partial_{\phi}. \end{aligned} \eeq

Writing $\{ {\bf X}_I\}_{I=1}^3 = \{ {\bf X}_x, {\bf X}_y, {\bf X}_z\}$ the non-zero commutator constants, $C^I_{~JK}$ are
\beq C^{{3}}_{~{1} {2}} = -1,~ C^{{2}}_{~{1} {3}} = 1,~C^{{1}}_{~{2} {3}} = -1 \eeq

The basis for the isotropy group is chosen to be:
\beq \lambda_{\hat{1}} = \left[ \begin{array}{cccc} 0 & 0 & 0 & 0 \\
0 & 0 & 0 & 0 \\ 0 & 0 & 0 & 1 \\ 0 & 0 & -1 & 0 \end{array} \right],~ \lambda_{\hat{2}} = -\left[ \begin{array}{cccc} 0 & 0 & 0 & 0 \\
0 & 0 & 1 & 0 \\ 0 & -1 & 0 & 0 \\ 0 & 0 & 0 & 0 \end{array} \right] ,~
\lambda_{\hat{3}} =  -\left[ \begin{array}{cccc} 0 & 0 & 0 & 0 \\
0 & 0 & 0 & 1 \\ 0 & 0 & 0 & 0 \\ 0 & -1 & 0 & 0 \end{array} \right]  \label{SSrep} \eeq

\noindent with the corresponding commutator constants, 
\beq C^{\hat{3}}_{~\hat{1} \hat{2}} = -1,~ C^{\hat{2}}_{~\hat{1} \hat{3}} = 1,~C^{\hat{1}}_{~\hat{2} \hat{3}} = -1 \eeq

To determine the group, $\overline{Iso}$ for this choice of isotropy generators, we again consider an arbitrary element of the Lie algebra of Lorentz frame transformations, $\tilde{\lambda}$ and impose that $[\lambda_{\hat{i}}, \tilde{\lambda}] \in ~\text{span}\{ \lambda_{\hat{1}}, \lambda_{\hat{2}},\lambda_{\hat{3}}\}$. The result of this calculation implies that $\tilde{\lambda} \in ~\text{span}\{ \lambda_{\hat{1}}, \lambda_{\hat{2}},\lambda_{\hat{3}}\}$ and so exponentiating each of the generators $\lambda_{\hat{i}},~\hat{i}=1,2,3$ shows that $\overline{Iso}$ consists of all spatial rotations. 

We will exploit the freedom of choice in the components of $f_I^{~\hat{i}}(x^{\alpha'})$ using the isotropy group, where the isotropy group affects a change to these components through equation \eqref{frot}. Since ${\bf X}_3 = {\bf X}_z$ is a generator of a spatial rotation, we will choose our frame so that ${\bf X}_3$ acts as a rotation on the basis elements $\bh^3$ and $\bh^4$, i.e., 
\beq f_{3}^{~\hat{i}} \lambda_{\hat{i}} = \left[ \begin{array}{cccc} 0 & 0 & 0 & 0 \\ 0 & 0 & 0 & 0 \\  0 & 0 & 0 & f_3^{~\hat{1}}  \\ 0 & 0 & -f_3^{~\hat{1}} & 0  \end{array} \right], \nonumber \eeq

\noindent then by applying a rotation about $\bh^3$ and $\bh^4$, the remaining component, $f_3^{~\hat{1}}$, can be set to zero using equation \eqref{frot}: 
\beq {\bf X}_3(\tilde{\Lambda}^a_{~b}) [\tilde{\Lambda}^{-1}]^b_{~c} + \tilde{\Lambda}^a_{~b} f_3^{~\hat{i}} \lambda^b_{\hat{i}~e} [\tilde{\Lambda}^{-1}]^e_{~c} = 0  \eeq

Choosing $f_3^{~\hat{i}}$ in this way, we can consider the combinations: ${\bf X}_1$ with ${\bf X}_3$, and ${\bf X}_2$ with ${\bf X}_3$ in equation \eqref{TP:sym3}. These combinations yield the following non-trivial equations:
\beq & \begin{aligned} & \partial_{\phi} f_1^{~\hat{1}} + f_2^{~\hat{1}} = 0, \partial_{\phi} f_1^{~\hat{2}} + f_2^{~\hat{2}} = 0, \partial_{\phi} f_1^{~\hat{3}} + f_2^{~\hat{3}} = 0. \\
& \partial_{\phi} f_2^{~\hat{1}} - f_1^{~\hat{1}} = 0, \partial_{\phi} f_2^{~\hat{2}} - f_1^{~\hat{2}} = 0, \partial_{\phi} f_2^{~\hat{3}} - f_1^{~\hat{3}} = 0, \end{aligned} \eeq 

\noindent which leads to the following solutions, 
\beq \begin{aligned} & f_2^{~\hat{i}} = f_2^{~\hat{i}}(\theta) \cos \phi + g_2^{~\hat{i}}(\theta) \sin \phi , \\
& f_1^{~\hat{i}} = \partial_{\phi} f_2^{~\hat{i}}. \end{aligned} \eeq

If we use Lorentz transformations that are dependent on $\theta$ alone, i.e., $\partial_{\phi} \tilde{\Lambda}^a_{~b} = 0$, then we can, without loss of generality, apply a rotation to set the coefficients of $\cos \phi$ in $f_2^{~\hat{i}}$ to zero using equation \eqref{frot}:
\beq {\bf X}_2(\tilde{\Lambda}^a_{~b}) [\tilde{\Lambda}^{-1}]^b_{~c} + \tilde{\Lambda}^a_{~b} f_2^{~\hat{i}} \lambda^b_{\hat{i}~e} [\tilde{\Lambda}^{-1}]^e_{~c} = -f_2^{~\hat{i}} \lambda^a_{~\hat{i}~c} \cos \phi. \eeq

\noindent Doing so, equation \eqref{TP:sym3} with ${\bf X}_2$ and ${\bf X}_1$ gives:
\beq \partial_{\theta} g_2^{~\hat{i}} + \tan(\theta) g_2^{~\hat{i}} = 0, \eeq

\noindent which has a simple solution:
\beq g_2^{~\hat{i}} = \frac{C_2^{~\hat{i}}}{\sin(\theta)}. \eeq

This fully determines $f_1^{~\hat{i}}$ and $f_2^{~\hat{i}}$, as the first is the $\phi$-derivative of the other.  Applying a rotation with constant parameters, we can set $f_2^{~\hat{2}}= f_2^{~\hat{3}} = 0$, without loss of generality we will choose the first component to be non-zero and set the arbitrary constant to one, and the solutions for the remaining components are:

\beq \begin{aligned}  f_2^{~\hat{1}} =  \frac{\sin \phi}{\sin \theta},~ f_1^{~\hat{1}} =  \frac{\cos \phi}{\sin \theta}. \end{aligned} \eeq

\noindent This gives the following matrix representation of the coefficients:
\beq f_I^{~\hat{i}} = \left [ \begin{array}{ccc}  \frac{\cos(\phi)}{\sin(\theta)} & 0 & 0 \\ \frac{\sin(\phi)}{\sin(\theta)} & 0 & 0 \\ 0 & 0 & 0 \end{array} \right] \label{SS:fmat}.\eeq

By making this choice, we have effectively fixed the frame up to $\overline{H}_q$. To determine this subgroup of $\overline{Iso}$, we will determine the form of $\tilde{\Lambda}$ which satisfy equation \eqref{frot} with $f_I^{~\hat{i}} = \tilde{f}_I^{~\hat{i}}$ taking the form of equation \eqref{SS:fmat}. We have the following three matrix equations: 

\beq \begin{aligned} & {\bf X}_1 ( \tilde{\Lambda}^a_{~b}) [\tilde{\Lambda}^{-1}]^b_{~c} + f_1^{~\hat{1}} [ \tilde{\Lambda}^a_{~b} \lambda^b_{\hat{1}~d} [\tilde{\Lambda}^{-1}]^d_{~c} -\lambda^a_{\hat{1}~ c} ] = 0 \\ 
& {\bf X}_2 ( \tilde{\Lambda}^a_{~b}) [\tilde{\Lambda}^{-1}]^b_{~c} + f_2^{~\hat{1}} [ \tilde{\Lambda}^a_{~b}\lambda^b_{\hat{1}~d} [\tilde{\Lambda}^{-1}]^d_{~c} - \lambda^a_{\hat{1}~ c} ] =0 \\
& {\bf X}_3 ( \tilde{\Lambda}^a_{~b}) [\tilde{\Lambda}^{-1}]^b_{~c} = 0. \end{aligned} \eeq

\noindent The first equation implies the parameters of $\tilde{\Lambda}^a_{~b}$ must be independent of $\phi$. Furthermore, using the chain rule, the first term can rewritten as, ${\bf X}_I ( \tilde{\Lambda}^a_{~b}) [\tilde{\Lambda}^{-1}]^b_{~c} = {\bf X}_I( g^{\hat{i}})) \lambda^a_{\hat{i}~c}$ where $g^{\hat{i}}$ are the parameters of $\tilde{\Lambda}^a_{~b}$. Lastly, we can write \beq  \tilde{\Lambda}^a_{~b}\lambda^b_{\hat{1}~d} [\tilde{\Lambda}^{-1}]^d_{~c}  = \alpha^{\hat{j}} \lambda^a_{\hat{j}~c}, \nonumber \eeq 
\noindent where the $\alpha^{\hat{j}}$ are expressions in terms of the parameters $g^{\hat{i}}$. The remaining equations can be summarized in a single equation:

\beq \begin{aligned} & \sin\phi \partial_{\theta} g^{\hat{i}}\lambda^a_{\hat{i}~c} - \frac{\cos\phi}{\sin \theta} [ \alpha^{\hat{j}} \lambda^a_{\hat{j}~c} -\lambda^a_{\hat{1}~ c} ] = 0. \\ 
\end{aligned} \eeq

\noindent The only solution, up to rescaling, of this differential equation is $g^{\hat{1}} = 1$ and   $g^{\hat{2}} = g^{\hat{3}} = 0$. This implies that the subgroup $\overline{H}_q$ consists of rotations in the $\bh^3-\bh^4$ plane with constant parameters.

Relative to the representation for the isotropy group in equation \eqref{SSrep}, we can solve the first equation in \eqref{Sym:RC:Prop}, 

\beq  { X}_I^{~\nu} \partial_{\nu} h^a_{~\mu} + \partial_{\mu} { X}_I^{~\nu} h^a_{~\nu} = f_I^{~\hat{i}} \lambda^a_{\hat{i}~b} h^b_{~\mu} \eeq 

\noindent  to determine the most general frame admitting this symmetry group:

\beq h^a_{~\mu} = \left[ \begin{array}{cccc} A_1(t,r) & A_2(t,r) & 0 & 0 \\ A_3(t,r) & A_4(t,r) & 0 & 0 \\ 0 & 0 & A_5(t,r) & -A_6(t,r) \sin(\theta) \\ 
0 & 0 & A_6(t,r) & A_5(t,r) \sin(\theta) \end{array}\right]. \label{VB:SS0} \eeq

\noindent We can choose a new coordinate system where the top block is diagonal and apply a spatial rotation to diagonalize the second block. The effect of these choices will be absorbed into the connection, which can be determined relative to this frame. The resulting vielbein is:

\beq h^a_{~\mu} = \left[ \begin{array}{cccc} A_1(t,r) & 0 & 0 & 0 \\ 0 & A_2(t,r) & 0 & 0 \\ 0 & 0 & A_3(t,r) & 0 \\ 
0 & 0 & 0 & A_3(t,r) \sin(\theta) \end{array}\right]. \label{VB:SS} \eeq

As the form of the $f_I^{~\hat{i}}$ in equation \eqref{SS:fmat} has been defined invariantly, the frame in equation \eqref{VB:SS0} is invariantly defined up to a linear isotropy subgroup $\overline{H}_q$ consisting of rotations in the $\bh^3-\bh^4$ plane with constant parameters. Relative to this coordinate system, the linear isotropy group can be further restricted by imposing a particular form on the vielbein as in equation \eqref{VB:SS}. With this choice the frame in equation \eqref{VB:SS} is an invariant symmetry frame, since the linear isotropy group is trivial.

With the chosen coframe, we can take the third equation in \eqref{Sym:RC:Prop} and solve for the most general connection. This is essentially a large set of differential equations for the connection coefficients and can be solved in a straightforward, albeit tedious manner.

\begin{prop}

The most general Riemann-Cartan geometry admitting spherical symmetry is given by the vielbein \eqref{VB:SS} and the metric compatible connection with components,

\beq \begin{aligned} & \omega_{341} = W_1(t,r),~ \omega_{342} = W_2(t,r), \\ 
& \omega_{233} = \omega_{244} = W_3(t,r),~ \omega_{234} = -\omega_{243} = W_4(t,r), \\
& \omega_{121} = W_5(t,r),~ \omega_{122} = W_6(t,r), \\
& \omega_{133} = \omega_{144} = W_7(t,r),~\omega_{134} = -\omega_{143} = W_8(t,r), \\
& \omega_{344} = - \frac{\cos(\theta)}{A_3 \sin(\theta)}. \end{aligned} \label{Con:SS} \eeq
\end{prop}

To determine the most general connection for a teleparallel geometry we next impose the flatness condition in \eqref{Sym:Prop}. The resulting equations are large but determining the solution is straightforward albeit tedious.

\begin{prop}
Any spherically symmetric teleparallel geometries is defined by the three arbitrary functions in the vielbein \eqref{VB:SS} along with the following spin connection components. 

\beq \begin{aligned} 
W_1 &= -\frac{\partial_t \chi}{A_1},W_2 = -\frac{\partial_r \chi}{A_2},\\
W_3 &= \frac{\cosh(\psi)\cos(\chi)}{A_3}, W_4 = \frac{\cosh(\psi)\sin(\chi)}{A_3},\\
W_5 &= -\frac{\partial_t \psi}{A_1}, W_6 = -\frac{\partial_r \psi}{A_2},\\
W_7 &= \frac{\sinh(\psi) \cos(\chi)}{A_3}, W_8 = \frac{\sinh(\psi) \sin(\chi)}{A_3},  
\end{aligned} \label{SS:TPcon} \eeq

\noindent where $\chi$ and $\psi$ are arbitrary functions of the coordinates $t$ and $r$.

\end{prop}

Any choice of the arbitrary functions $\psi$ and $\chi$, picks out a unique teleparallel geometry, as any change in the form of the spin connection which could affect the form of $\psi$ or $\chi$ leads to a change in the form of the vielbein. For a given pair of functions, the invariantly defined frame up to the linear isotropy group $H_q$ arising from the CK algorithm could be computed to provide further sub-classification. We note that there are only five arbitrary functions required to specify a geometry: $A_1, A_2, A_3, \psi$ and $\chi$. This characterization of  spherically symmetric geometries in terms of $\psi$ and $\chi$ has not been presented before in the literature and provides a formulation of spherically symmetric teleparallel geometries using the smallest number of arbitrary functions. With this, we can determine new families of geometries by imposing conditions on the arbitrary functions where these conditions are related to  the inertial effects of the geometries. 

While these geometries have been investigated in the  teleparallel context before using the proper frame approach \cite{hohmann2019modified,pfeifer2022quick,pfeifer2021static}, a thorough investigation in the context of Riemann-Cartan geometries with non-vanishing torsion has not been performed. We note that subclasses of these  geometries have been studied earlier in teleparallel gravity \cite{sharif2009teleparallel} using the Killing equations for an arbitrary spherically symmetric metric. However, such an approach does not consider the sensitivity of the frame basis to Lorentz transformations and does not pick out the most general classes of spherically symmetric teleparallel geometries as the parameters of the Lorentz transformations do not appear at all in their formulation. 

In the context of teleparallel gravity, our approach determines the minimal number of arbitrary functions, at most five, in the vielbein and spin connection to define the geometries. It also readily distinguishes four inequivalent branches of spherically symmetric teleparallel geometries, depending on whether $\chi$ or $\psi$ vanish, as the frame in \eqref{VB:SS} has been fully fixed by restricting the vielbein to a particular form. In comparison, the proper frame approach requires at most six arbitrary functions and the inequivalence of two spherically symmetric teleparallel geometries will require the full machinery of the CK algorithm \cite{hohmann2019modified, pfeifer2022quick}.

We will conclude this section with a brief discussion of stationary spherically symmetric geometries. To the authors' knowledge, very little is known about the stationary spherically symmetric Riemann-Cartan geometries with non-vanishing torsion. Stationary or static spherically symmetric teleparallel geometries have been discussed in applications \cite{sharif2009teleparallel, bahamonde2021exploring, pfeifer2021static}. In these papers, the stationarity or static condition is not rigorously derived but instead arises from restricting the arbitrary functions to be dependent on the radial coordinate alone. Although, such a choice intuitively makes sense, it does not necessarily constitute the most general class of stationary or static spherically symmetric geometries since it does not consider the inequivalent subclasses of spherically symmetric metrics.
 
In a similar manner to the axisymmetric geometries, stationary spherically symmetric solutions can be produced by including an additional affine frame symmetry, ${\bf Y}$ with $Y^a Y_a < 0$, locally. This vector field must commute with the original frame symmetries, $[{\bf X}_I, {\bf Y}] = 0$ for $I \in \{1, \ldots, 3\}$. Therefore, a new coordinate system can be chosen so that ${\bf Y} = \partial_{t}$, while preserving the vielbein form in \eqref{VB:SS0} instead of the more specialized vielbein in \eqref{VB:SS}. The timelike condition on ${\bf Y}$ imposes an additional condition on the vielbein functions $A_i$, $i\in \{1, \ldots, 6\}$ in equation \eqref{VB:SS0}. Static spherically symmetric geometries are produced by requiring that ${\bf Y}$ is hypersurface orthogonal:  
\beq {\bf Y} \wedge d {\bf Y} = 0. \eeq
\noindent In the case of static solutions, the vielbein form \eqref{VB:SS} is applicable. The expansion of the symmetry group introduces three additional  functions into \eqref{TP:sym1} and this can be determined using \eqref{TP:sym3}. As the commutator constants are zero, the analysis of these equations is straightforward and imply that the vielbein and the components of the connection one-form must be dependent on the coordinate $r$ alone. As no other conditions are imposed, this proves that the static spherically symmetric solutions can be found by requiring that the functions $A_i,~i=1,2,3$, $\psi$ and $\chi$ are functions of the $r$ coordinate. 

We emphasize that we have not attempted to address the anti-symmetric field equations for $F(T)$ theories. While this has been considered previously using the proper frame approach \cite{pfeifer2022quick} for particular choices of the arbitrary functions, a general analysis of the field equations has not been carried out \cite{Bahamonde:2021gfp}. However, the phenomenology of spherically symmetric static and stationary solutions in $F(T)$ theories have been explored in \cite{bahamonde2022black,bahamonde2019photon, debenedictis2016spherically,bahamonde2020solar}. In our approach, the minimal number of arbitrary functions for each inequivalent subclass can potentially allow for a general analysis of the field equations in $F(T)$ theories.

\subsection{Robertson-Walker geometries}

Working in the coordinate system  $(t, r, \theta, \phi)$, the affine frame symmetry generators associated with the RW metric are

\beq \begin{aligned} & {\bf X}_1 = \zeta \sin \theta \cos \phi \partial_r + \frac{\zeta}{r}\cos \theta \cos \phi \partial_{\theta} - \frac{\zeta \sin \phi}{r \sin \theta} \partial_{\phi}, \\
& {\bf X}_2 = \zeta \sin \theta \sin \phi \partial_r + \frac{\zeta}{r} \cos \theta \sin \phi \partial_{\theta} + \frac{\zeta \cos \phi}{r \sin \theta} \partial_{\phi}, \\ 
& {\bf X}_3 = \zeta \cos \theta \partial_r - \frac{\zeta}{r} \sin \theta \partial_{\theta}, \\ & {\bf X}_z = \partial_{\phi},~{\bf X}_y = - \cos \phi \partial_{\theta} + \frac{\sin \phi}{\tan \theta} \partial_{\phi},~ {\bf X}_x = \sin \phi \partial_{\theta} + \frac{\cos \phi}{\tan \theta} \partial_{\phi}, \end{aligned} \eeq

\noindent where $\zeta = \sqrt{1-kr^2}$ and $k \in \mathbb{R}$. Writing $\{ {\bf X}_I\}_{I=1}^6 = \{ {\bf X}_1, {\bf X}_2, {\bf X}_3, {\bf X}_x, {\bf X}_y, {\bf X}_z\}$ the non-zero commutator constants, $C^I_{~JK}$ are

\beq \begin{aligned} &  C^6_{~12} = -k, C^5_{~13} = -k, C^3_{~15} = 1 \\ 
& C^2_{~16} = 1,~ C^4_{~23} = k,~ C^3_{~24} = -1, \\
& C^1_{~26} = -1,~ C^2_{~34} = 1,~ C^1_{~35} = -1, \\
& C^6_{~45} = -1,~ C^5_{~46} = 1,~ C^4_{~56} = -1. \end{aligned} \eeq

The basis for the  isotropy group is of the form

\beq \lambda_{\hat{1}} = \left[ \begin{array}{cccc} 0 & 0 & 0 & 0 \\
0 & 0 & 0 & 0 \\ 0 & 0 & 0 & 1 \\ 0 & 0 & -1 & 0 \end{array} \right],~ \lambda_{\hat{2}} = -\left[ \begin{array}{cccc} 0 & 0 & 0 & 0 \\
0 & 0 & 1 & 0 \\ 0 & -1 & 0 & 0 \\ 0 & 0 & 0 & 0 \end{array} \right] ,~
\lambda_{\hat{3}} =  -\left[ \begin{array}{cccc} 0 & 0 & 0 & 0 \\
0 & 0 & 0 & 1 \\ 0 & 0 & 0 & 0 \\ 0 & -1 & 0 & 0 \end{array} \right]  \label{Isochoice} \eeq

\noindent with the corresponding commutator constants, 
\beq C^{\hat{3}}_{~\hat{1} \hat{2}} = -1,~ C^{\hat{2}}_{~\hat{1} \hat{3}} = 1,~C^{\hat{1}}_{~\hat{2} \hat{3}} = -1. \eeq

\noindent The form of $\overline{Iso}$ can be determined in the same manner as in case of spherical symmetry since the isotropy group is unchanged. Thus, $\overline{Iso}$ consists of spatial rotations. 

With these quantities, we can substitute them into the first two equations in \eqref{Sym:RC:Prop}, first by solving the second equation to determine the form of the functions in $f_I^{~\hat{i}}$. 
\beq 2{\bf X}_{[I} ( f_{J]}^{~\hat{k}}) - f_I^{~\hat{i}} f_J^{~\hat{j}} C^{\hat{k}}_{~\hat{i} \hat{j}} = C^K_{~IJ} f_K^{~\hat{k}}. \label{FLRWstart} \eeq 

\noindent We will exploit the freedom of choice in the components of $f_I^{~\hat{i}}$ using the  isotropy group to transform the components with equation \eqref{frot}. Since ${\bf X}_z$ is a generator of a spatial rotation, we will choose our frame representation so that it acts as a rotation on the basis elements $\bh^3$ and $\bh^4$, i.e., 
\beq f_{6}^{~\hat{i}} \lambda_{\hat{i}} = \left[ \begin{array}{cccc} 0 & 0 & 0 & 0 \\ 0 & 0 & 0 & 0 \\  0 & 0 & 0 & f_6^{~\hat{1}}  \\ 0 & 0 & -f_6^{~\hat{1}} & 0  \end{array} \right], \nonumber \eeq

\noindent then by applying a rotation about $\bh^3$ and $\bh^4$, the remaining component, $f_6^{~\hat{1}}$, can be set to zero using
\beq {\bf X}_6(\tilde{\Lambda}^a_{~b}) [\tilde{\Lambda}^{-1}]^b_{~c} + \tilde{\Lambda}^a_{~b} f_6^{~\hat{i}} \lambda^b_{\hat{i}~e} [\tilde{\Lambda}^{-1}]^e_{~c} = 0  \eeq
From the second equation in \eqref{Sym:RC:Prop}, using ${\bf X}_6$ paired with ${\bf X}_{\overline{I}}$, $\overline{I} \in [1,5]$, we have the following differential equations:

\beq & \begin{aligned} & \partial_{\phi} f_1^{~\hat{1}} + f_2^{~\hat{1}} = 0, \partial_{\phi} f_1^{~\hat{2}} + f_2^{~\hat{2}} = 0, \partial_{\phi} f_1^{~\hat{3}} + f_3^{~\hat{1}} = 0, \\
& \partial_{\phi} f_2^{~\hat{1}} + f_1^{~\hat{1}} = 0, \partial_{\phi} f_2^{~\hat{2}} + f_1^{~\hat{2}} = 0, \partial_{\phi} f_2^{~\hat{3}} + f_1^{~\hat{3}} = 0, \\
& \partial_{\phi} f_3^{~\hat{1}} = 0, \partial_{\phi} f_3^{~\hat{2}} = 0, \partial_{\phi} f_3^{~\hat{3}} = 0, \\
& \partial_{\phi} f_4^{~\hat{1}} + f_5^{~\hat{1}} = 0, \partial_{\phi} f_4^{~\hat{2}} + f_5^{~\hat{2}} = 0, \partial_{\phi} f_4^{~\hat{3}} + f_5^{~\hat{3}} = 0 \\
& \partial_{\phi} f_5^{~\hat{1}} + f_4^{~\hat{1}} = 0, \partial_{\phi} f_5^{~\hat{2}} + f_4^{~\hat{2}} = 0, \partial_{\phi} f_5^{~\hat{3}} + f_4^{~\hat{3}} = 0 \end{aligned} \eeq 

\noindent Leading to the following solutions, 

\beq \begin{aligned} & f_5^{~\hat{i}} = f_5^{~\hat{i}}(r,\theta) \cos \phi + g_5^{~\hat{i}}(r,\theta) \sin \phi , \\
& f_4^{~\hat{i}} = \partial_{\phi} f_5^{~\hat{i}}, \\
& f_3^{~\hat{i}} = f_3^{~\hat{i}}(r,\theta), \\
& f_2^{~\hat{i}} = f_2^{~\hat{i}}(r,\theta) \cos \phi + g_2^{~\hat{i}}(r,\theta) \sin \phi,\\ 
& f_1^{~\hat{i}} = \partial_{\phi} f_2^{~\hat{i}}.\end{aligned} \eeq

If Lorentz transformations are restricted to transformations that are dependent on $r$ and $\theta$, we can apply a rotation to set the coefficients of $\cos \phi$ in $f_5^{~\hat{i}}$ to zero using equation \eqref{frot} which takes the form:
\beq {\bf X}_5(\tilde{\Lambda}^a_{~b}) [\tilde{\Lambda}^{-1}]^b_{~c} + \tilde{\Lambda}^a_{~b} f_5^{~\hat{i}} \lambda^b_{\hat{i}~e} [\tilde{\Lambda}^{-1}]^e_{~c} = -f_5^{~\hat{i}} \lambda^a_{~\hat{i}~c} \cos \phi. \eeq

\noindent Doing so, the second equation in \eqref{Sym:RC:Prop} with ${\bf X}_5$ and ${\bf X}_4$ gives
\beq \partial_{\theta} g_5^{~\hat{i}} + \cos(\theta) g_5^{~\hat{i}} = 0, \eeq

\noindent with the resulting solution:
\beq g_5^{~\hat{i}} = \frac{G_5^{~\hat{i}}(r)}{\sin(\theta)}. \eeq

This fully determines $f_5^{~\hat{i}}$ and $f_4^{~\hat{i}}$ up to functions of $r$, as the second is the $\phi$-derivative of the first. Applying a rotation dependent on only the coordinates $r$ we can fix some of the components in $f_5^{~\hat{i}}$ to be zero. Without loss of generality, we will choose the first component to be non-zero:
\beq \begin{aligned} & f_5^{~\hat{1}} =  \frac{\sin \phi}{\sin \theta}, \\ & f_4^{~\hat{1}} =  \frac{\cos \phi}{\sin \theta}. \end{aligned} \eeq

The remaining equations in \eqref{FLRWstart} can now be solved by looking at the coefficients of $\cos \phi$, and $\sin \phi$. With these choices, $f_I^{~\hat{i}}$ takes the form:

\beq f_I^{~\hat{i}} = \left [ \begin{array}{ccc} - \frac{\sqrt{1-kr^2} \sin(\phi) \cos(\theta)}{r \sin(\theta)} & -\frac{\cos(\theta) \cos(\phi)}{r} & \frac{\sin(\phi)}{r} \\ \frac{\sqrt{1-kr^2} \cos(\phi) \cos(\theta)}{r \sin(\theta)} & -\frac{\cos(\theta) \sin(\phi)}{r} & - \frac{\cos(\phi)}{r} \\ 0 & \frac{\sin(\theta)}{r} & 0 \\ \frac{\cos(\phi)}{\sin(\theta)} & 0 & 0 \\ \frac{\sin(\phi)}{\sin(\theta)} & 0 & 0 \\ 0 & 0 & 0 \end{array} \right] \label{TRWfmatrix}\eeq

\noindent The form of $f_I^{~\hat{i}}$ determines the parameters of Lorentz transformations in $\overline{Iso}$ up to $\overline{H}_q$. 

To determine $\overline{H}_q$, we employ equation \eqref{frot} with $f_I^{~\hat{i}}  = \tilde{f}_I^{~\hat{i}}$ defined as in equation \eqref{TRWfmatrix}. The analysis proceeds in the same manner as the spherically symmetric case, except that the parameters of $\tilde{\Lambda}^a_{~b}$ now depend on $r, \theta$ and $\phi$. Considering the Lie algebra generator of $\tilde{\Lambda}^a_{~b}$ as $\tilde{\lambda}^a_{~c} = g^{\hat{i}} \lambda^a_{\hat{i}~b}$, then using ${\bf X}_4, {\bf X}_5$ and ${\bf X}_6$, equation \eqref{frot} implies that $ g^{\hat{1}} = g(r)$ and $g^{\hat{2}}= g^{\hat{3}} = 0$. Finally equation \eqref{frot} with ${\bf X}_3$ gives

\beq \zeta \cos \theta \partial_r f \lambda^a_{\hat{1}~c} + \frac{\sin \theta}{r} [ (\cos f) \lambda^a_{\hat{2}~c} + (\sin f) \lambda^a_{\hat{3}~c} - \lambda^a_{\hat{2}~c}] = 0 \eeq

\noindent The only solution to this equation is $f=0$, implying that $\tilde{\Lambda}^a_{~b}$ is the identity Lorentz transformation. Thus, the subgroup $\overline{H}_q$ is the trivial subgroup. 

Using the representation of the  isotropy group in \eqref{Isochoice}, and setting the components of $f_I^{~\hat{i}}$ into the first equation in \eqref{Sym:RC:Prop} yields the following general solution for the vielbein: 

\beq h^a_{~\mu} = \left[ \begin{array}{cccc} A_1(t) & 0 & 0 & 0 \\ 0 & \frac{A_2(t)}{\sqrt{1-kr^2}} & 0 & 0 \\ 0 & 0 & A_2(t) r & 0 \\ 0 & 0 & 0 & A_2(t) r \sin(\theta) \end{array} \right]. \label{TRWvielbein} \eeq

\noindent Imposing the vielbein form in equation \eqref{TRWvielbein}, since $\overline{H}_q$ is the trivial group, this frame is now an invariant frame.  In general, a coordinate transformation can be made to set $A_1 = 1$ and $A_2 = a(t)$. Doing so, we will work with the coframe associated with 

\beq h^a_{~\mu} = \left[ \begin{array}{cccc} 1 & 0 & 0 & 0 \\ 0 & \frac{a(t)}{\sqrt{1-kr^2}} & 0 & 0 \\ 0 & 0 & a(t) r & 0 \\ 0 & 0 & 0 & a(t) r \sin(\theta) \end{array} \right]. \label{VB:FLRW} \eeq 

Using the coframe associated to this matrix, we can readily solve the resulting equations coming from equations \eqref{Liederivative:Con} for each of the affine frame symmetries. The resulting differential equations can be solved readily by hand or by using the frame and connection ansatz from the spherically symmetric case.

\begin{prop}

The most general Riemann-Cartan geometry which admits a RW symmetry group consists of the vielbein \eqref{VB:FLRW} along with the metric-compatible connection with components:
\beq \begin{aligned} & \omega_{122} = \omega_{133} = \omega_{144} =  W_1(t), \\
& \omega_{234} = -\omega_{243} = \omega_{342} = W_2(t), \\
& \omega_{233} = \omega_{244} = - \frac{\sqrt{1-kr^2}}{a(t)r}, \\
& \omega_{344} = - \frac{\cos(\theta)}{a(t) r \sin(\theta)}. \end{aligned} \label{Con:FLRW} \eeq

\end{prop}

The class of connections that describe a teleparallel geometry can be determined by solving the flatness condition in equation \eqref{Sym:Prop}. This gives the following differential constraints on $W_1$ and $W_2$:

\beq \begin{aligned} & a \partial_t W_1 + W_1 \partial_t a  = 0, \\
& W_1 W_2 = 0, \\ 
& a \partial_t W_2 + W_2 \partial_t a  = 0, \\
& W_2^2 a^2 - W_1^2 a^2 - k \end{aligned} \eeq

\noindent These equations have four distinct solutions, leading to the following proposition

\begin{prop}
The most general teleparallel geometry admitting the RW symmetry group are specified by the arbitrary function in the vielbein \eqref{VB:FLRW} along with the choice of a metric-compatible connection \eqref{Con:FLRW} with the functions $W_1$ and $W_2$ given as
\begin{enumerate}
\item $W_1 = 0$ and $W_2 = - \frac{\sqrt{k}}{a}$. 
\item $W_1 = 0$ and $W_2 =  \frac{\sqrt{k}}{a}$. 
\item $W_1 = - \frac{\sqrt{-k}}{a}$ and $W_2 = 0$.  
\item $W_1 = \frac{\sqrt{-k}}{a}$ and $W_2 = 0$. 
\end{enumerate}

\noindent Each choice of $W_1$ and $W_2$ yields an inequivalent teleparallel geometry.
\end{prop}

Imposing the reasonable condition that the torsion tensors must be real-valued, this naturally distinguishes two cases for $k>0$ and two cases for $k<0$. When $k$ vanishes, these four inequivalent cases coincide. Without loss of generality, the coordinate  $r$ can be rescaled to set $k = -1,0$ or $1$. Note that the antisymmetric field equations are automatically satisfied in all cases.  In \cite{Coley:2022}, the proper frames for these teleparallel geometries are given and the $F(T)$ field equations are analysed for each inequivalent case. 

Alternatively, the connection coefficients for this case can be determined using the connection coefficients of the spherically symmetric case in equation \eqref{SS:TPcon} by setting 
\beq \begin{aligned} k<0 &: \chi = 0, \text{or } \pi,~\psi = \cosh^{-1} (-\cos(\chi) \sqrt{1-kr^2}), \\
k \geq 0 &: \chi =  \cos^{-1} (- \sqrt{1-kr^2}), \psi = 0. \end{aligned} \eeq
\noindent This choice of representation of the arbitrary functions in the connection coefficients shows that the sign difference in the case of $k >0$ is purely a choice of representation and hence these subcases are equivalent. We note that the sign difference for the $k<0$ case does indeed give two inequivalent subcases. 

While the RW Riemann-Cartan geometries with non-vanishing torsion have not been explored at all, the teleparallel RW geometries have been discussed previously. Using a metric ansatz and the Killing equations a very limited subclass were determined in \cite{sharif2009teleparallel}, and the flat case has been explored by many authors \cite{cai2016f}. The first attempt at using a frame approach to determine TRW geometries with $k \neq 0$ were in \cite{Ferraro:2006jd,Ferraro:2008ey}. However, the solutions required a complex orthonormal frame which was undesirable from a physical standpoint.

The general class was determined using the proper frame approach in \cite{hohmann2021complete}. We note that the field equations studied in \cite{Coley:2022} agree with the field equations found in \cite{hohmann2021complete}. However, there are two significant problems with the approach in \cite{hohmann2021complete}: certain branches of solutions require complex orthonormal proper frames \cite{pfeifer2022quick}, and the proper frames contain an additional arbitrary function of $t$ which is purely a coordinate artefact.  Despite the appearance of complex orthonormal proper frames, these have been used to study the RW models with non-zero $k$ in \cite{hohmann2021teleparallel, hohmann2021general}. 

Many of the applications in the literature of these geometries have concentrated on the analysis of the flat ($k = 0$) TRW cosmological models \cite{Bahamonde:2021gfp, cai2016f}. In addition, particular forms for $F(T)$ have been considered. For example, employing reconstruction methods, a functional form for the solution, in terms of the scale factor, is assumed and the corresponding $F(T)$ theory is then constructed. However, geometries with non-zero $k$ have been recently studied in bounce models \cite{casalino2021bounce} and in inflation where the use of a complex orthonormal frame potentially restricts to the $k=1$ case \cite{capozziello2018cosmic}. Lastly, the TRW solutions have been used to study perturbations in non-flat cosmology \cite{bahamonde2022perturbations}. More generally, the TRW solutions with non-zero $k$ have been employed in $f(T,B)$ theories, where $B$ is a boundary term \cite{paliathanasis2022f}.

\subsection{de Sitter and Einstein Static geometries  }

In GR and Lorentzian geometry, in general, the de Sitter solution is a special case of the RW solutions, where the six-dimensional group is expanded to a ten-dimensional group, namely $SO(1,4)$. It has been proven that the only maximally symmetric geometry which admits a non-vanishing torsion tensor is Minkowski space and that the largest symmetry group permitted in such a geometry is at most seven \cite{Coley:2019zld, Bahamonde:2021gfp}. In addition, it follows from the CK algorithm that if a geometry admits a seven-dimensional symmetry group then it must have constant Cartan invariants relative to some class of invariantly defined frames \cite{Coley:2019zld}. 

To determine the possible special cases of the TRW geometries which admit admit a seven-dimensional symmetry group, we will first consider the general case of the Riemann-Cartan geometries and require that their components are constant. The components of the curvature tensor and the torsion tensor are, respectively, 
\beq \begin{aligned} &R_{1212} = R_{1313} = R_{1414} = \frac{W_{2,t} a + a_{,t} W_2}{a},\\ 
&R_{1234} = R_{1324} = R_{1423} = 2 W_1 W_2, \\
&R_{2314} = R_{2413} = R_{3412} =  \frac{W_{1,t} a + a_{,t} W_1}{a},\\
&R_{2323} = R_{2424} = R_{3434}= \frac{W_1^2 a^2 - W_2^2 a^2 - k}{a^2}, \end{aligned} \eeq
\noindent and 
\beq \begin{aligned}
& T_{212} = T_{313} = T_{414} = \frac{W_2 a + a_{,t}}{a},\\ 
& T_{234} = T_{324} = T_{432} = 2 W_1.  \end{aligned} \eeq

\noindent Requiring that these components are constant give two distinct cases. 

\begin{prop}

The subclass of RW Riemann-Cartan geometries admitting a seventh symmetry are given by the vielbein \eqref{VB:FLRW} and the connection \eqref{Con:FLRW} where the functions $W_1$, $W_2$ and $a(t)$ take the form:

\beq \begin{aligned} & \text{ Case 1: } W_1 = C_1, W_2 = C_2, k=0, \text{ and } a(t) = C_3 e^{C_4 t}, \\
& \text{ Case 2: } W_1 = C_1, W_2 = C_2, \text{ k arbitrary, and } a(t) = C_3. 
 \end{aligned} \eeq
 
 \end{prop}
 
Requiring the curvature tensor vanishes leads to strong conditions on $W_1$ and $W_2$. In the remainder of this subsection, we will only discuss the teleparallel geometries admitting a 7-dimensional symmetry group, $G_7$. 

\begin{prop}
The subclass of RW teleparallel geometries admitting a seventh symmetry are given by the vielbein \eqref{VB:FLRW} and the connection \eqref{Con:FLRW} with

\begin{enumerate}
\item $W_1 = 0$ and $W_2 = - \frac{\sqrt{k}}{a}$. 
\item $W_1 = 0$ and $W_2 =  \frac{\sqrt{k}}{a}$. 
\item $W_1 = - \frac{\sqrt{-k}}{a}$ and $W_2 = 0$.  
\item $W_1 = \frac{\sqrt{-k}}{a}$ and $W_2 = 0$. 
\end{enumerate}

\noindent where either 

\beq \begin{aligned} &  \text{ Case 1: } a = C_3 e^{C_4t} \text{ and } k = 0, \\
& \text{ Case 2: } a = C_3 \text{ and k is $\pm 1$}. \end{aligned} \eeq

\end{prop}

Using \eqref{TP:frm:sym} and \eqref{Liederivative:Con} we can determine the form of the seventh symmetry as 
\beq \begin{aligned} 
& \text{ Case 1: } {\bf X}_7 = \frac{1}{C_3 C_4} \partial_t + r \partial_r, \\
& \text{ Case 2: } {\bf X}_7 =  \partial_t. \end{aligned} \eeq

\noindent In the first case, the resulting Lie algebra of $\{{\bf X}_I\}_{I=1}^{7} = \{{\bf X}_1,{\bf X}_2,{\bf X}_3, {\bf X}_x, {\bf X}_y, {\bf X}_z, {\bf X}_7\}$ is given by the following non-zero Lie brackets:

\begin{equation}
\begin{array}{lll}
{}  [{\bf X}_1,{\bf X}_5]={\bf X}_3,  & [{\bf X}_1,{\bf X}_6]={\bf X}_2,  & [{\bf X}_1,{\bf X}_7]={\bf X}_1, \\
{} [{\bf X}_2,{\bf X}_4]=-{\bf X}_3, & [{\bf X}_2,{\bf X}_6]=-{\bf X}_1, & [{\bf X}_2,{\bf X}_7]={\bf X}_2, \\
{}  [{\bf X}_3,{\bf X}_4]={\bf X}_2,  & [{\bf X}_3,{\bf X}_5]=-{\bf X}_1, & [{\bf X}_3,{\bf X}_7]={\bf X}_3, \\
{}  [{\bf X}_4,{\bf X}_5]=-{\bf X}_6, & [{\bf X}_4,{\bf X}_6]={\bf X}_5,  & [{\bf X}_5,{\bf X}_6]=-{\bf X}_4.
\end{array} \nonumber
\end{equation}
By inspection, this is a subalgebra of the Lie algebra for the group of metric (Killing) symmetries of de Sitter spacetime.
We therefore propose the following definition.
\begin{defn}
The teleparallel de Sitter geometry (TdS) is a teleparallel geometry with the $G_7$ Lie group of affine symmetries, $\mathcal{R} \rtimes E(3)$ which is a subgroup of $O(1,4)$.
\end{defn}
\noindent We note in this geometry the covariant derivative of the torsion tensor is zero.

In the second case, $a(t) = C_3$ a constant, $k=\pm1$.  These geometries correspond to the direct product $\mathbb{R} \times M_3$ where $M_3$ is a locally homogeneous and isotropic Riemannian manifold. This is reflected in the Lie algebra structure of the affine frame symmetries  $\{{\bf X}_I\}_{I=1}^{7} = \{{\bf X}_1,{\bf X}_2,{\bf X}_3, {\bf X}_x, {\bf X}_y, {\bf X}_z, {\bf X}_7\}$ where ${\bf X}_7 = \partial_t$ satisfies $[{\bf X}_i, {\bf X}_7] = 0, i \in [1,6]$ and the remaining non-zero Lie brackets are:

\begin{equation}
 \begin{array}{lll}
{} [{\bf X}_1,{\bf X}_5]={\bf X}_3, & [{\bf X}_1,{\bf X}_6]={\bf X}_2,& [{\bf X}_2,{\bf X}_4]=-{\bf X}_3, \\
{} [{\bf X}_2,{\bf X}_6]=-{\bf X}_1,& [{\bf X}_3,{\bf X}_4]={\bf X}_2,& [{\bf X}_3,{\bf X}_5]=-{\bf X}_1, \\
{} [{\bf X}_4,{\bf X}_5]=-{\bf X}_6,& [{\bf X}_4,{\bf X}_6]={\bf X}_5,& [{\bf X}_5,{\bf X}_6]=-{\bf X}_4. \end{array} \nonumber
\end{equation}
Since ${\bf X}_7 = \partial_t$ is the additional affine frame symmetry, this geometry is necessarily static due to the form of the frame basis.  This geometry can be considered as the analogue of the Einstein static geometry in GR which we call the Teleparallel Einstein Static (TES) geometry. We therefore propose the following definition.

\begin{defn}
The teleparallel Einstein Static geometry (TES) is a teleparallel geometry with the $G_7$ Lie group of affine symmetries, $\mathcal{R} \times E(3)$ which is a subgroup of $O(1,4)$.
\end{defn}

By working with the RW geometries, which are a subclass of the spherically symmetric geometries, admitting six affine frame symmetries we have determined two subclasses of Riemann-Cartan geometries which admit two distinct seven-dimensional groups of affine frame symmetries, respectively. In the case of teleparallel geometries, we have explicitly determined the analogues of the de Sitter (TdS) and Einstein static (TES) geometries. We note that the proper frames and the $F(T)$ field equations for the TdS and TES geometries were presented in \cite{ColeyvdH2022}. These are only two examples of geometries admitting a seven-dimensional group of affine frame symmetries, and we anticipate a much larger set of geometries admitting a six-dimensional group of affine frame symmetries with different six-dimensional groups. In the case of geometries admitting a seven-dimensional group of affine frame symmetries, these will necessarily belong to the class of locally homogeneous Riemann-Cartan geometries \cite{McNutt:2022}.

\section{Discussion} \label{sec:Discussion}

We have introduced an invariant definition for symmetries in frame based theories of gravity. This is a generalization of the Killing equations in pseudo-Riemannian geometry to the more general class of Riemann-Cartan geometries. As an illustration of our approach, we have provided examples from teleparallel geometries where the connection is chosen so that the torsion tensor and its covariant derivatives are the only non-vanishing geometric quantities. This definition relies on a geometrically preferred class of frames, called {\it symmetry frames}, which are a generalization of the class of invariantly defined frames, or invariant frames used in the Cartan-Karlhede (CK) algorithm. Relative to the symmetry frames, the equations for an affine frame symmetry \eqref{Intro2} are put into a simple form given in equations \eqref{TP:sym1} and \eqref{TP:sym2} which can be solved readily for a coframe $\bh^a$ connection one-form $\bomega^a_{~b}$ and Lie algebra generator of the isotropy group, $\lambda^a_{~b}$. In general, an invariantly defined frame can then be constructed by fixing the form of $f_I^{~\hat{i}}$ in equation \eqref{TP:sym1} in a coordinate independent manner. Furthermore, an invariant frame, where all parameters of the Lorentz transformations are fixed, can be found by restricting the form of the vielbein. This permits the inequivalent subclasses of geometries to be readily distinguished.

In addition to this definition, we have introduced an approach to determine the most general Riemann-Cartan geometries which admit a specified symmetry group. Once a particular coordinate system and representation of the symmetry group as vector fields are given, the equations for an affine frame symmetry, \eqref{TP:sym1} and \eqref{TP:sym2}, along with the affine frame symmetry commutator condition \eqref{TP:sym3} can be solved for the coframe $\bh^a$ and connection one-form $\bomega^a_{~b}$. In order to restrict to teleparallel geometries, additional conditions must be imposed on the connection one-form, namely the vanishing of the curvature tensor. We show that for the connection one-form that satisfies \eqref{TP:sym2}, imposing the flatness condition can be achieved by solving a system of differential equations. 

The geometrical approach in the proper frame approach \cite{pfeifer2022quick, hohmann2019modified, hohmann2021complete} is similar to that given here. The definitions are not identical, and we are careful only to discuss affine symmetries and avoid any complex quantities. Since we have made no assumptions a priori, our results are general. A significant difference between the approaches is that, in general, the invariantly defined symmetry frame formulation requires a non-vanishing spin connection. While this may complicate the presentation of the geometries which admit a given symmetry group, the inclusion of the spin connection permits a complete classification of solutions of equations \eqref{TP:sym1} and \eqref{TP:sym2}.

While we have determined two distinct classes of Riemann-Cartan geometries admitting two particular seven-dimensional groups of symmetries, it is known that that there are other Riemann-Cartan geometries which admit a group of seven affine frame symmetries. Within the framework of the symmetry frame approach, it is possible to explicitly find all possible Riemann-Cartan geometries admitting a seven dimensional symmetry group, and in future work we will present an exhaustive list of these. We will also investigate the class of geometries which permit a six-dimensional group of symmetries in order to determine the teleparallel analogue of the locally homogeneous pp-waves. The class of all Riemann-Cartan geometries admitting six-dimensional groups of affine frame symmetries are more  difficult to determine as these can permit two-dimensional linear isotropy groups in addition to three-dimensional linear isotropy groups.

We note that the approach discussed in this paper can be extended to connections that are not metric compatible and hence admit a non-vanishing non-metricity tensor $Q_{abc} = \nabla_{\bh_c} g_{ab}$ and more general groups of frame transformations such as $SO(p,q)$. The authors believe that this can provide a helpful tool for exploring more exotic geometrics or more general pseudo-Riemannian geometries with indefinite signature and give an insight into the alignment classification \cite{hervik2015degenerate}.

\section*{Acknowledgments}
AAC and RvdH are supported by the Natural Sciences and Engineering Research Council of Canada. RvdH is supported by the St Francis Xavier University Council on Research. DDM is supported by the Norwegian Financial Mechanism 2014-2021 (project registration number 2019/34/H/ST1/00636). The authors thank Alexandre Landry for comments and discussion.


\bibliographystyle{apsrev4-2}
\bibliography{Tele-Parallel-Reference-file0}

\begin{thebibliography}{41}%
\makeatletter
\providecommand \@ifxundefined [1]{%
 \@ifx{#1\undefined}
}%
\providecommand \@ifnum [1]{%
 \ifnum #1\expandafter \@firstoftwo
 \else \expandafter \@secondoftwo
 \fi
}%
\providecommand \@ifx [1]{%
 \ifx #1\expandafter \@firstoftwo
 \else \expandafter \@secondoftwo
 \fi
}%
\providecommand \natexlab [1]{#1}%
\providecommand \enquote  [1]{``#1''}%
\providecommand \bibnamefont  [1]{#1}%
\providecommand \bibfnamefont [1]{#1}%
\providecommand \citenamefont [1]{#1}%
\providecommand \href@noop [0]{\@secondoftwo}%
\providecommand \href [0]{\begingroup \@sanitize@url \@href}%
\providecommand \@href[1]{\@@startlink{#1}\@@href}%
\providecommand \@@href[1]{\endgroup#1\@@endlink}%
\providecommand \@sanitize@url [0]{\catcode `\\12\catcode `\$12\catcode
  `\&12\catcode `\#12\catcode `\^12\catcode `\_12\catcode `\%12\relax}%
\providecommand \@@startlink[1]{}%
\providecommand \@@endlink[0]{}%
\providecommand \url  [0]{\begingroup\@sanitize@url \@url }%
\providecommand \@url [1]{\endgroup\@href {#1}{\urlprefix }}%
\providecommand \urlprefix  [0]{URL }%
\providecommand \Eprint [0]{\href }%
\providecommand \doibase [0]{https://doi.org/}%
\providecommand \selectlanguage [0]{\@gobble}%
\providecommand \bibinfo  [0]{\@secondoftwo}%
\providecommand \bibfield  [0]{\@secondoftwo}%
\providecommand \translation [1]{[#1]}%
\providecommand \BibitemOpen [0]{}%
\providecommand \bibitemStop [0]{}%
\providecommand \bibitemNoStop [0]{.\EOS\space}%
\providecommand \EOS [0]{\spacefactor3000\relax}%
\providecommand \BibitemShut  [1]{\csname bibitem#1\endcsname}%
\let\auto@bib@innerbib\@empty
\bibitem [{\citenamefont {Chinea}(1988)}]{chinea1988symmetries}%
  \BibitemOpen
  \bibfield  {author} {\bibinfo {author} {\bibfnamefont {F.}~\bibnamefont
  {Chinea}},\ }\href@noop {} {\bibfield  {journal} {\bibinfo  {journal}
  {Classical and Quantum Gravity}\ }\textbf {\bibinfo {volume} {5}},\ \bibinfo
  {pages} {135} (\bibinfo {year} {1988})}\BibitemShut {NoStop}%
\bibitem [{\citenamefont {Estabrook}\ and\ \citenamefont
  {Wahlquist}(1996)}]{estabrook1996moving}%
  \BibitemOpen
  \bibfield  {author} {\bibinfo {author} {\bibfnamefont {F.~B.}\ \bibnamefont
  {Estabrook}}\ and\ \bibinfo {author} {\bibfnamefont {H.~D.}\ \bibnamefont
  {Wahlquist}},\ }\href@noop {} {\bibfield  {journal} {\bibinfo  {journal}
  {Classical and Quantum Gravity}\ }\textbf {\bibinfo {volume} {13}},\ \bibinfo
  {pages} {1333} (\bibinfo {year} {1996})}\BibitemShut {NoStop}%
\bibitem [{\citenamefont {Papadopoulos}\ and\ \citenamefont
  {Grammenos}(2012)}]{papadopoulos2012locally}%
  \BibitemOpen
  \bibfield  {author} {\bibinfo {author} {\bibfnamefont {G.~O.}\ \bibnamefont
  {Papadopoulos}}\ and\ \bibinfo {author} {\bibfnamefont {T.}~\bibnamefont
  {Grammenos}},\ }\href@noop {} {\bibfield  {journal} {\bibinfo  {journal}
  {Journal of Mathematical Physics}\ }\textbf {\bibinfo {volume} {53}},\
  \bibinfo {pages} {072502} (\bibinfo {year} {2012})}\BibitemShut {NoStop}%
\bibitem [{\citenamefont {Olver}(1995)}]{olver1995equivalence}%
  \BibitemOpen
  \bibfield  {author} {\bibinfo {author} {\bibfnamefont {P.~J.}\ \bibnamefont
  {Olver}},\ }\href@noop {} {\emph {\bibinfo {title} {Equivalence, invariants
  and symmetry}}}\ (\bibinfo  {publisher} {Cambridge University Press},\
  \bibinfo {year} {1995})\BibitemShut {NoStop}%
\bibitem [{\citenamefont {Aldrovandi}\ and\ \citenamefont
  {Pereira}(2013)}]{Aldrovandi_Pereira2013}%
  \BibitemOpen
  \bibfield  {author} {\bibinfo {author} {\bibfnamefont {R.}~\bibnamefont
  {Aldrovandi}}\ and\ \bibinfo {author} {\bibfnamefont {J.~G.}\ \bibnamefont
  {Pereira}},\ }\href {https://doi.org/10.1007/978-94-007-5143-9} {\emph
  {\bibinfo {title} {{Teleparallel Gravity}}}},\ \bibinfo {series} {Fundamental
  Theories of Physics}, Vol.\ \bibinfo {volume} {173}\ (\bibinfo  {publisher}
  {Springer},\ \bibinfo {address} {Dordrecht},\ \bibinfo {year}
  {2013})\BibitemShut {NoStop}%
\bibitem [{\citenamefont {Ferraro}\ and\ \citenamefont
  {Fiorini}(2007)}]{Ferraro:2006jd}%
  \BibitemOpen
  \bibfield  {author} {\bibinfo {author} {\bibfnamefont {R.}~\bibnamefont
  {Ferraro}}\ and\ \bibinfo {author} {\bibfnamefont {F.}~\bibnamefont
  {Fiorini}},\ }\href {https://doi.org/10.1103/PhysRevD.75.084031} {\bibfield
  {journal} {\bibinfo  {journal} {Physical Review}\ }\textbf {\bibinfo {volume}
  {D75}},\ \bibinfo {pages} {084031} (\bibinfo {year} {2007})},\ \Eprint
  {https://arxiv.org/abs/gr-qc/0610067} {arXiv:gr-qc/0610067 [gr-qc]}
  \BibitemShut {NoStop}%
\bibitem [{\citenamefont {Ferraro}\ and\ \citenamefont
  {Fiorini}(2008)}]{Ferraro:2008ey}%
  \BibitemOpen
  \bibfield  {author} {\bibinfo {author} {\bibfnamefont {R.}~\bibnamefont
  {Ferraro}}\ and\ \bibinfo {author} {\bibfnamefont {F.}~\bibnamefont
  {Fiorini}},\ }\href {https://doi.org/10.1103/PhysRevD.78.124019} {\bibfield
  {journal} {\bibinfo  {journal} {Physical Review}\ }\textbf {\bibinfo {volume}
  {D78}},\ \bibinfo {pages} {124019} (\bibinfo {year} {2008})},\ \Eprint
  {https://arxiv.org/abs/0812.1981} {arXiv:0812.1981 [gr-qc]} \BibitemShut
  {NoStop}%
\bibitem [{\citenamefont {Linder}(2010)}]{Linder:2010py}%
  \BibitemOpen
  \bibfield  {author} {\bibinfo {author} {\bibfnamefont {E.~V.}\ \bibnamefont
  {Linder}},\ }\href {https://doi.org/10.1103/PhysRevD.81.127301,
  10.1103/PhysRevD.82.109902} {\bibfield  {journal} {\bibinfo  {journal}
  {Physical Review}\ }\textbf {\bibinfo {volume} {D81}},\ \bibinfo {pages}
  {127301} (\bibinfo {year} {2010})},\ \bibinfo {note} {[Erratum: Phys.
  Rev.D82,109902(2010)]},\ \Eprint {https://arxiv.org/abs/1005.3039}
  {arXiv:1005.3039 [astro-ph.CO]} \BibitemShut {NoStop}%
\bibitem [{\citenamefont {Kr\v{s}\v{s}\'{a}k}\ \emph
  {et~al.}(2019)\citenamefont {Kr\v{s}\v{s}\'{a}k}, \citenamefont {van~den
  Hoogen}, \citenamefont {Pereira}, \citenamefont {Boehmer},\ and\
  \citenamefont {Coley}}]{Krssak:2018ywd}%
  \BibitemOpen
  \bibfield  {author} {\bibinfo {author} {\bibfnamefont {M.}~\bibnamefont
  {Kr\v{s}\v{s}\'{a}k}, \bibfnamefont {M.}}, \bibinfo {author} {\bibfnamefont
  {R.~J.}\ \bibnamefont {van~den Hoogen}}, \bibinfo {author} {\bibfnamefont
  {J.~G.}\ \bibnamefont {Pereira}}, \bibinfo {author} {\bibfnamefont {C.~G.}\
  \bibnamefont {Boehmer}},\ and\ \bibinfo {author} {\bibfnamefont {A.~A.}\
  \bibnamefont {Coley}},\ }\href {https://doi.org/10.1088/1361-6382/ab2e1f}
  {\bibfield  {journal} {\bibinfo  {journal} {Classical and Quantum Gravity}\
  }\textbf {\bibinfo {volume} {36}},\ \bibinfo {pages} {183001} (\bibinfo
  {year} {2019})},\ \Eprint {https://arxiv.org/abs/1810.12932}
  {arXiv:1810.12932 [gr-qc]} \BibitemShut {NoStop}%
\bibitem [{\citenamefont {Lucas}\ \emph {et~al.}(2009)\citenamefont {Lucas},
  \citenamefont {Obukhov},\ and\ \citenamefont
  {Pereira}}]{Lucas_Obukhov_Pereira2009}%
  \BibitemOpen
  \bibfield  {author} {\bibinfo {author} {\bibfnamefont {T.~G.}\ \bibnamefont
  {Lucas}}, \bibinfo {author} {\bibfnamefont {Y.~N.}\ \bibnamefont {Obukhov}},\
  and\ \bibinfo {author} {\bibfnamefont {J.~G.}\ \bibnamefont {Pereira}},\
  }\href {https://doi.org/10.1103/PhysRevD.80.064043} {\bibfield  {journal}
  {\bibinfo  {journal} {Physical Review}\ }\textbf {\bibinfo {volume} {D80}},\
  \bibinfo {pages} {064043} (\bibinfo {year} {2009})},\ \Eprint
  {https://arxiv.org/abs/0909.2418} {arXiv:0909.2418 [gr-qc]} \BibitemShut
  {NoStop}%
\bibitem [{\citenamefont {Kr\v{s}\v{s}\'{a}k}\ and\ \citenamefont
  {Pereira}(2015)}]{Krssak_Pereira2015}%
  \BibitemOpen
  \bibfield  {author} {\bibinfo {author} {\bibfnamefont {M.}~\bibnamefont
  {Kr\v{s}\v{s}\'{a}k}}\ and\ \bibinfo {author} {\bibfnamefont {J.~G.}\
  \bibnamefont {Pereira}},\ }\href
  {https://doi.org/10.1140/epjc/s10052-015-3749-2} {\bibfield  {journal}
  {\bibinfo  {journal} {The European Physical Journal}\ }\textbf {\bibinfo
  {volume} {C75}},\ \bibinfo {pages} {519} (\bibinfo {year} {2015})},\ \Eprint
  {https://arxiv.org/abs/1504.07683} {arXiv:1504.07683 [gr-qc]} \BibitemShut
  {NoStop}%
\bibitem [{\citenamefont {Pfeifer}(2022)}]{pfeifer2022quick}%
  \BibitemOpen
  \bibfield  {author} {\bibinfo {author} {\bibfnamefont {C.}~\bibnamefont
  {Pfeifer}},\ }\href@noop {} {\bibfield  {journal} {\bibinfo  {journal} {A
  quick guide to spacetime symmetry and symmetric solutions in teleparallel
  gravity}\ } (\bibinfo {year} {2022})},\ \Eprint
  {https://arxiv.org/abs/2201.04691} {arXiv:2201.04691 [gr-qc]} \BibitemShut
  {NoStop}%
\bibitem [{\citenamefont {Hohmann}\ \emph {et~al.}(2019)\citenamefont
  {Hohmann}, \citenamefont {J{\"a}rv}, \citenamefont {Kr{\v{s}}{\v{s}}{\'a}k},\
  and\ \citenamefont {Pfeifer}}]{hohmann2019modified}%
  \BibitemOpen
  \bibfield  {author} {\bibinfo {author} {\bibfnamefont {M.}~\bibnamefont
  {Hohmann}}, \bibinfo {author} {\bibfnamefont {L.}~\bibnamefont {J{\"a}rv}},
  \bibinfo {author} {\bibfnamefont {M.}~\bibnamefont
  {Kr{\v{s}}{\v{s}}{\'a}k}},\ and\ \bibinfo {author} {\bibfnamefont
  {C.}~\bibnamefont {Pfeifer}},\ }\href@noop {} {\bibfield  {journal} {\bibinfo
   {journal} {Physical Review}\ }\textbf {\bibinfo {volume} {100}},\ \bibinfo
  {pages} {084002} (\bibinfo {year} {2019})}\BibitemShut {NoStop}%
\bibitem [{\citenamefont {Hohmann}(2021{\natexlab{a}})}]{hohmann2021complete}%
  \BibitemOpen
  \bibfield  {author} {\bibinfo {author} {\bibfnamefont {M.}~\bibnamefont
  {Hohmann}},\ }\href@noop {} {\bibfield  {journal} {\bibinfo  {journal}
  {International Journal of Geometric Methods in Mathematical Physics}\
  }\textbf {\bibinfo {volume} {18}},\ \bibinfo {pages} {2140005} (\bibinfo
  {year} {2021}{\natexlab{a}})}\BibitemShut {NoStop}%
\bibitem [{\citenamefont {Coley}\ \emph {et~al.}(2020)\citenamefont {Coley},
  \citenamefont {Van Den~Hoogen},\ and\ \citenamefont
  {McNutt}}]{Coley:2019zld}%
  \BibitemOpen
  \bibfield  {author} {\bibinfo {author} {\bibfnamefont {A.~A.}\ \bibnamefont
  {Coley}}, \bibinfo {author} {\bibfnamefont {R.~J.}\ \bibnamefont {Van
  Den~Hoogen}},\ and\ \bibinfo {author} {\bibfnamefont {D.~D.}\ \bibnamefont
  {McNutt}},\ }\href {https://doi.org/10.1063/5.0003252} {\bibfield  {journal}
  {\bibinfo  {journal} {Journal of Mathematical Physics}\ }\textbf {\bibinfo
  {volume} {61}},\ \bibinfo {pages} {072503} (\bibinfo {year} {2020})},\
  \Eprint {https://arxiv.org/abs/1911.03893} {arXiv:1911.03893 [gr-qc]}
  \BibitemShut {NoStop}%
\bibitem [{\citenamefont {Bahamonde}\ \emph {et~al.}(2023)\citenamefont
  {Bahamonde}, \citenamefont {Dialektopoulos}, \citenamefont
  {Escamilla-Rivera}, \citenamefont {Farrugia}, \citenamefont {Gakis},
  \citenamefont {Hendry}, \citenamefont {Hohmann}, \citenamefont {Said},
  \citenamefont {Mifsud},\ and\ \citenamefont
  {Di~Valentino}}]{Bahamonde:2021gfp}%
  \BibitemOpen
  \bibfield  {author} {\bibinfo {author} {\bibfnamefont {S.}~\bibnamefont
  {Bahamonde}}, \bibinfo {author} {\bibfnamefont {K.~F.}\ \bibnamefont
  {Dialektopoulos}}, \bibinfo {author} {\bibfnamefont {C.}~\bibnamefont
  {Escamilla-Rivera}}, \bibinfo {author} {\bibfnamefont {G.}~\bibnamefont
  {Farrugia}}, \bibinfo {author} {\bibfnamefont {V.}~\bibnamefont {Gakis}},
  \bibinfo {author} {\bibfnamefont {M.}~\bibnamefont {Hendry}}, \bibinfo
  {author} {\bibfnamefont {M.}~\bibnamefont {Hohmann}}, \bibinfo {author}
  {\bibfnamefont {J.~L.}\ \bibnamefont {Said}}, \bibinfo {author}
  {\bibfnamefont {J.}~\bibnamefont {Mifsud}},\ and\ \bibinfo {author}
  {\bibfnamefont {E.}~\bibnamefont {Di~Valentino}},\ }\href@noop {} {\bibfield
  {journal} {\bibinfo  {journal} {Reports on Progress in Physics}\ }\textbf
  {\bibinfo {volume} {86}},\ \bibinfo {pages} {026901} (\bibinfo {year}
  {2023})},\ \Eprint {https://arxiv.org/abs/2106.13793} {arXiv:2106.13793
  [gr-qc]} \BibitemShut {NoStop}%
\bibitem [{\citenamefont {Fonseca-Neto}\ \emph {et~al.}(1996)\citenamefont
  {Fonseca-Neto}, \citenamefont {Reboucas},\ and\ \citenamefont
  {MacCallum}}]{fonseca1996algebraic}%
  \BibitemOpen
  \bibfield  {author} {\bibinfo {author} {\bibfnamefont {J.~B.}\ \bibnamefont
  {Fonseca-Neto}}, \bibinfo {author} {\bibfnamefont {M.~J.}\ \bibnamefont
  {Reboucas}},\ and\ \bibinfo {author} {\bibfnamefont {M.~A.~H.}\ \bibnamefont
  {MacCallum}},\ }\href@noop {} {\bibfield  {journal} {\bibinfo  {journal}
  {Mathematics and Computers in Simulation}\ }\textbf {\bibinfo {volume}
  {42}},\ \bibinfo {pages} {739} (\bibinfo {year} {1996})}\BibitemShut
  {NoStop}%
\bibitem [{\citenamefont {{\AA}man}\ \emph {et~al.}(1998)\citenamefont
  {{\AA}man}, \citenamefont {Fonseca-Neto}, \citenamefont {MacCallum},\ and\
  \citenamefont {Rebou{\c{c}}as}}]{aaman1998riemann}%
  \BibitemOpen
  \bibfield  {author} {\bibinfo {author} {\bibfnamefont {J.~E.}\ \bibnamefont
  {{\AA}man}}, \bibinfo {author} {\bibfnamefont {J.~B.}\ \bibnamefont
  {Fonseca-Neto}}, \bibinfo {author} {\bibfnamefont {M.~A.~H.}\ \bibnamefont
  {MacCallum}},\ and\ \bibinfo {author} {\bibfnamefont {M.~J.}\ \bibnamefont
  {Rebou{\c{c}}as}},\ }\href@noop {} {\bibfield  {journal} {\bibinfo  {journal}
  {Classical and Quantum Gravity}\ }\textbf {\bibinfo {volume} {15}},\ \bibinfo
  {pages} {1089} (\bibinfo {year} {1998})}\BibitemShut {NoStop}%
\bibitem [{\citenamefont {Swift}\ \emph {et~al.}(1986)\citenamefont {Swift},
  \citenamefont {d'Inverno},\ and\ \citenamefont
  {Vickers}}]{swift1986everywhere}%
  \BibitemOpen
  \bibfield  {author} {\bibinfo {author} {\bibfnamefont {S.~T.}\ \bibnamefont
  {Swift}}, \bibinfo {author} {\bibfnamefont {R.~A.}\ \bibnamefont
  {d'Inverno}},\ and\ \bibinfo {author} {\bibfnamefont {J.~A.~G.}\ \bibnamefont
  {Vickers}},\ }\href@noop {} {\bibfield  {journal} {\bibinfo  {journal}
  {General Relativity and Gravitation}\ }\textbf {\bibinfo {volume} {18}},\
  \bibinfo {pages} {1093} (\bibinfo {year} {1986})}\BibitemShut {NoStop}%
\bibitem [{Note1()}]{Note1}%
  \BibitemOpen
  \bibinfo {note} {A null, or even complex null frame will work as well,
  depending on the desired representation of the isotropy group}\BibitemShut
  {NoStop}%
\bibitem [{\citenamefont {Yano}(2020)}]{yano2020theory}%
  \BibitemOpen
  \bibfield  {author} {\bibinfo {author} {\bibfnamefont {K.}~\bibnamefont
  {Yano}},\ }\href@noop {} {\emph {\bibinfo {title} {The theory of Lie
  derivatives and its applications}}}\ (\bibinfo  {publisher} {Courier Dover
  Publications},\ \bibinfo {year} {2020})\BibitemShut {NoStop}%
\bibitem [{\citenamefont {Hall}(2004)}]{hall2004symmetries}%
  \BibitemOpen
  \bibfield  {author} {\bibinfo {author} {\bibfnamefont {G.~S.}\ \bibnamefont
  {Hall}},\ }\href@noop {} {\emph {\bibinfo {title} {Symmetries and curvature
  structure in general relativity}}},\ Vol.~\bibinfo {volume} {46}\ (\bibinfo
  {publisher} {world scientific},\ \bibinfo {year} {2004})\BibitemShut
  {NoStop}%
\bibitem [{\citenamefont {Nashed}(2010)}]{nashed2010stationary}%
  \BibitemOpen
  \bibfield  {author} {\bibinfo {author} {\bibfnamefont {G.~G.}\ \bibnamefont
  {Nashed}},\ }\href@noop {} {\bibfield  {journal} {\bibinfo  {journal}
  {Astrophysics and Space Science}\ }\textbf {\bibinfo {volume} {330}},\
  \bibinfo {pages} {173} (\bibinfo {year} {2010})}\BibitemShut {NoStop}%
\bibitem [{\citenamefont {Pfeifer}\ and\ \citenamefont
  {Schuster}(2021)}]{pfeifer2021static}%
  \BibitemOpen
  \bibfield  {author} {\bibinfo {author} {\bibfnamefont {C.}~\bibnamefont
  {Pfeifer}}\ and\ \bibinfo {author} {\bibfnamefont {S.}~\bibnamefont
  {Schuster}},\ }\href@noop {} {\bibfield  {journal} {\bibinfo  {journal} {MDPI
  Universe}\ }\textbf {\bibinfo {volume} {7}},\ \bibinfo {pages} {153}
  (\bibinfo {year} {2021})}\BibitemShut {NoStop}%
\bibitem [{\citenamefont {Sharif}\ and\ \citenamefont
  {Majeed}(2009)}]{sharif2009teleparallel}%
  \BibitemOpen
  \bibfield  {author} {\bibinfo {author} {\bibfnamefont {M.}~\bibnamefont
  {Sharif}}\ and\ \bibinfo {author} {\bibfnamefont {B.}~\bibnamefont
  {Majeed}},\ }\href@noop {} {\bibfield  {journal} {\bibinfo  {journal}
  {Communications in Theoretical Physics}\ }\textbf {\bibinfo {volume} {52}},\
  \bibinfo {pages} {435} (\bibinfo {year} {2009})}\BibitemShut {NoStop}%
\bibitem [{\citenamefont {Bahamonde}\ \emph {et~al.}(2021)\citenamefont
  {Bahamonde}, \citenamefont {Valcarcel}, \citenamefont {J{\"a}rv},\ and\
  \citenamefont {Pfeifer}}]{bahamonde2021exploring}%
  \BibitemOpen
  \bibfield  {author} {\bibinfo {author} {\bibfnamefont {S.}~\bibnamefont
  {Bahamonde}}, \bibinfo {author} {\bibfnamefont {J.~G.}\ \bibnamefont
  {Valcarcel}}, \bibinfo {author} {\bibfnamefont {L.}~\bibnamefont
  {J{\"a}rv}},\ and\ \bibinfo {author} {\bibfnamefont {C.}~\bibnamefont
  {Pfeifer}},\ }\href@noop {} {\bibfield  {journal} {\bibinfo  {journal}
  {Physical Review D}\ }\textbf {\bibinfo {volume} {103}},\ \bibinfo {pages}
  {044058} (\bibinfo {year} {2021})}\BibitemShut {NoStop}%
\bibitem [{\citenamefont {Bahamonde}\ \emph
  {et~al.}(2022{\natexlab{a}})\citenamefont {Bahamonde}, \citenamefont
  {Golovnev}, \citenamefont {Guzm{\'a}n}, \citenamefont {Said},\ and\
  \citenamefont {Pfeifer}}]{bahamonde2022black}%
  \BibitemOpen
  \bibfield  {author} {\bibinfo {author} {\bibfnamefont {S.}~\bibnamefont
  {Bahamonde}}, \bibinfo {author} {\bibfnamefont {A.}~\bibnamefont {Golovnev}},
  \bibinfo {author} {\bibfnamefont {M.~J.}\ \bibnamefont {Guzm{\'a}n}},
  \bibinfo {author} {\bibfnamefont {J.~L.}\ \bibnamefont {Said}},\ and\
  \bibinfo {author} {\bibfnamefont {C.}~\bibnamefont {Pfeifer}},\ }\href@noop
  {} {\bibfield  {journal} {\bibinfo  {journal} {Journal of Cosmology and
  Astroparticle Physics}\ }\textbf {\bibinfo {volume} {2022}}\bibinfo  {number}
  { (01)},\ \bibinfo {pages} {037}}\BibitemShut {NoStop}%
\bibitem [{\citenamefont {Bahamonde}\ \emph {et~al.}(2019)\citenamefont
  {Bahamonde}, \citenamefont {Flathmann},\ and\ \citenamefont
  {Pfeifer}}]{bahamonde2019photon}%
  \BibitemOpen
\bibfield  {number} {  }\bibfield  {author} {\bibinfo {author} {\bibfnamefont
  {S.}~\bibnamefont {Bahamonde}}, \bibinfo {author} {\bibfnamefont
  {K.}~\bibnamefont {Flathmann}},\ and\ \bibinfo {author} {\bibfnamefont
  {C.}~\bibnamefont {Pfeifer}},\ }\href@noop {} {\bibfield  {journal} {\bibinfo
   {journal} {Physical Review D}\ }\textbf {\bibinfo {volume} {100}},\ \bibinfo
  {pages} {084064} (\bibinfo {year} {2019})}\BibitemShut {NoStop}%
\bibitem [{\citenamefont {DeBenedictis}\ and\ \citenamefont
  {Ilijic}(2016)}]{debenedictis2016spherically}%
  \BibitemOpen
  \bibfield  {author} {\bibinfo {author} {\bibfnamefont {A.}~\bibnamefont
  {DeBenedictis}}\ and\ \bibinfo {author} {\bibfnamefont {S.}~\bibnamefont
  {Ilijic}},\ }\href@noop {} {\bibfield  {journal} {\bibinfo  {journal} {{The
  Spherically Symmetric Vacuum in Covariant $ F (T)= T+ \frac{\alpha}{2} T^2 +
  O(T^{\gamma})$ Gravity Theory}}\ } (\bibinfo {year} {2016})},\ \Eprint
  {https://arxiv.org/abs/1609.07465} {arXiv:1609.07465 [gr-qc]} \BibitemShut
  {NoStop}%
\bibitem [{\citenamefont {Bahamonde}\ \emph {et~al.}(2020)\citenamefont
  {Bahamonde}, \citenamefont {Said},\ and\ \citenamefont
  {Zubair}}]{bahamonde2020solar}%
  \BibitemOpen
  \bibfield  {author} {\bibinfo {author} {\bibfnamefont {S.}~\bibnamefont
  {Bahamonde}}, \bibinfo {author} {\bibfnamefont {J.~L.}\ \bibnamefont
  {Said}},\ and\ \bibinfo {author} {\bibfnamefont {M.}~\bibnamefont {Zubair}},\
  }\href@noop {} {\bibfield  {journal} {\bibinfo  {journal} {Journal of
  Cosmology and Astroparticle Physics}\ }\textbf {\bibinfo {volume}
  {2020}}\bibinfo  {number} { (10)},\ \bibinfo {pages} {024}}\BibitemShut
  {NoStop}%
\bibitem [{\citenamefont {Coley}\ \emph {et~al.}(2022)\citenamefont {Coley},
  \citenamefont {Van Den~Hoogen},\ and\ \citenamefont {McNutt}}]{Coley:2022}%
  \BibitemOpen
\bibfield  {number} {  }\bibfield  {author} {\bibinfo {author} {\bibfnamefont
  {A.~A.}\ \bibnamefont {Coley}}, \bibinfo {author} {\bibfnamefont {R.~J.}\
  \bibnamefont {Van Den~Hoogen}},\ and\ \bibinfo {author} {\bibfnamefont
  {D.~D.}\ \bibnamefont {McNutt}},\ }\href@noop {} {\bibfield  {journal}
  {\bibinfo  {journal} {Classical and Quantum Gravity}\ }\textbf {\bibinfo
  {volume} {39}},\ \bibinfo {pages} {22LT01} (\bibinfo {year} {2022})},\
  \Eprint {https://arxiv.org/abs/2205.10719} {arXiv:2205.10719 [gr-qc]}
  \BibitemShut {NoStop}%
\bibitem [{\citenamefont {Cai}\ \emph {et~al.}(2016)\citenamefont {Cai},
  \citenamefont {Capozziello}, \citenamefont {De~Laurentis},\ and\
  \citenamefont {Saridakis}}]{cai2016f}%
  \BibitemOpen
  \bibfield  {author} {\bibinfo {author} {\bibfnamefont {Y.~F.}\ \bibnamefont
  {Cai}}, \bibinfo {author} {\bibfnamefont {S.}~\bibnamefont {Capozziello}},
  \bibinfo {author} {\bibfnamefont {M.}~\bibnamefont {De~Laurentis}},\ and\
  \bibinfo {author} {\bibfnamefont {E.~N.}\ \bibnamefont {Saridakis}},\
  }\href@noop {} {\bibfield  {journal} {\bibinfo  {journal} {Reports on
  Progress in Physics}\ }\textbf {\bibinfo {volume} {79}},\ \bibinfo {pages}
  {106901} (\bibinfo {year} {2016})}\BibitemShut {NoStop}%
\bibitem [{\citenamefont {Hohmann}\ and\ \citenamefont
  {Pfeifer}(2021)}]{hohmann2021teleparallel}%
  \BibitemOpen
  \bibfield  {author} {\bibinfo {author} {\bibfnamefont {M.}~\bibnamefont
  {Hohmann}}\ and\ \bibinfo {author} {\bibfnamefont {C.}~\bibnamefont
  {Pfeifer}},\ }\href@noop {} {\bibfield  {journal} {\bibinfo  {journal} {The
  European Physical Journal C}\ }\textbf {\bibinfo {volume} {81}},\ \bibinfo
  {pages} {1} (\bibinfo {year} {2021})}\BibitemShut {NoStop}%
\bibitem [{\citenamefont {Hohmann}(2021{\natexlab{b}})}]{hohmann2021general}%
  \BibitemOpen
  \bibfield  {author} {\bibinfo {author} {\bibfnamefont {M.}~\bibnamefont
  {Hohmann}},\ }\href@noop {} {\bibfield  {journal} {\bibinfo  {journal} {The
  European Physical Journal Plus}\ }\textbf {\bibinfo {volume} {136}},\
  \bibinfo {pages} {1} (\bibinfo {year} {2021}{\natexlab{b}})}\BibitemShut
  {NoStop}%
\bibitem [{\citenamefont {Casalino}\ \emph {et~al.}(2021)\citenamefont
  {Casalino}, \citenamefont {Sanna}, \citenamefont {Sebastiani},\ and\
  \citenamefont {Zerbini}}]{casalino2021bounce}%
  \BibitemOpen
  \bibfield  {author} {\bibinfo {author} {\bibfnamefont {A.}~\bibnamefont
  {Casalino}}, \bibinfo {author} {\bibfnamefont {B.}~\bibnamefont {Sanna}},
  \bibinfo {author} {\bibfnamefont {L.}~\bibnamefont {Sebastiani}},\ and\
  \bibinfo {author} {\bibfnamefont {S.}~\bibnamefont {Zerbini}},\ }\href@noop
  {} {\bibfield  {journal} {\bibinfo  {journal} {Physical Review D}\ }\textbf
  {\bibinfo {volume} {103}},\ \bibinfo {pages} {023514} (\bibinfo {year}
  {2021})}\BibitemShut {NoStop}%
\bibitem [{\citenamefont {Capozziello}\ \emph {et~al.}(2018)\citenamefont
  {Capozziello}, \citenamefont {Luongo}, \citenamefont {Pincak},\ and\
  \citenamefont {Ravanpak}}]{capozziello2018cosmic}%
  \BibitemOpen
  \bibfield  {author} {\bibinfo {author} {\bibfnamefont {S.}~\bibnamefont
  {Capozziello}}, \bibinfo {author} {\bibfnamefont {O.}~\bibnamefont {Luongo}},
  \bibinfo {author} {\bibfnamefont {R.}~\bibnamefont {Pincak}},\ and\ \bibinfo
  {author} {\bibfnamefont {A.}~\bibnamefont {Ravanpak}},\ }\href@noop {}
  {\bibfield  {journal} {\bibinfo  {journal} {General Relativity and
  Gravitation}\ }\textbf {\bibinfo {volume} {50}},\ \bibinfo {pages} {1}
  (\bibinfo {year} {2018})}\BibitemShut {NoStop}%
\bibitem [{\citenamefont {Bahamonde}\ \emph
  {et~al.}(2022{\natexlab{b}})\citenamefont {Bahamonde}, \citenamefont
  {Dialektopoulos}, \citenamefont {Hohmann}, \citenamefont {Said},
  \citenamefont {Pfeifer},\ and\ \citenamefont
  {Saridakis}}]{bahamonde2022perturbations}%
  \BibitemOpen
  \bibfield  {author} {\bibinfo {author} {\bibfnamefont {S.}~\bibnamefont
  {Bahamonde}}, \bibinfo {author} {\bibfnamefont {K.~F.}\ \bibnamefont
  {Dialektopoulos}}, \bibinfo {author} {\bibfnamefont {M.}~\bibnamefont
  {Hohmann}}, \bibinfo {author} {\bibfnamefont {J.~L.}\ \bibnamefont {Said}},
  \bibinfo {author} {\bibfnamefont {C.}~\bibnamefont {Pfeifer}},\ and\ \bibinfo
  {author} {\bibfnamefont {E.~N.}\ \bibnamefont {Saridakis}},\ }\href@noop {}
  {\bibfield  {journal} {\bibinfo  {journal} {Perturbations in Non-Flat
  Cosmology for $ f (T) $ gravity}\ } (\bibinfo {year} {2022}{\natexlab{b}})},\
  \Eprint {https://arxiv.org/abs/2203.00619} {arXiv:2203.00619 [gr-qc]}
  \BibitemShut {NoStop}%
\bibitem [{\citenamefont {Paliathanasis}\ and\ \citenamefont
  {Leon}(2022)}]{paliathanasis2022f}%
  \BibitemOpen
  \bibfield  {author} {\bibinfo {author} {\bibfnamefont {A.}~\bibnamefont
  {Paliathanasis}}\ and\ \bibinfo {author} {\bibfnamefont {G.}~\bibnamefont
  {Leon}},\ }\href@noop {} {\bibfield  {journal} {\bibinfo  {journal} {$ f (T,
  B) $ gravity in a Friedmann-Lemaitre-Robertson-Walker universe with nonzero
  spatial curvature}\ } (\bibinfo {year} {2022})},\ \Eprint
  {https://arxiv.org/abs/2201.12189} {arXiv:2201.12189 [gr-qc]} \BibitemShut
  {NoStop}%
\bibitem [{\citenamefont {Coley}\ and\ \citenamefont {van~den
  Hoogen}(2023)}]{ColeyvdH2022}%
  \BibitemOpen
  \bibfield  {author} {\bibinfo {author} {\bibfnamefont {A.}~\bibnamefont
  {Coley}}\ and\ \bibinfo {author} {\bibfnamefont {R.}~\bibnamefont {van~den
  Hoogen}},\ }\href {https://doi.org/10.1063/5.0099551} {\bibfield  {journal}
  {\bibinfo  {journal} {J. Math. Phys.}\ }\textbf {\bibinfo {volume} {64}},\
  \bibinfo {pages} {022503} (\bibinfo {year} {2023})},\ \Eprint
  {https://arxiv.org/abs/2205.07071} {arXiv:2205.07071 [gr-qc]} \BibitemShut
  {NoStop}%
\bibitem [{\citenamefont {McNutt}\ \emph {et~al.}(2022)\citenamefont {McNutt},
  \citenamefont {Coley},\ and\ \citenamefont {Van Den~Hoogen}}]{McNutt:2022}%
  \BibitemOpen
  \bibfield  {author} {\bibinfo {author} {\bibfnamefont {D.~D.}\ \bibnamefont
  {McNutt}}, \bibinfo {author} {\bibfnamefont {A.~A.}\ \bibnamefont {Coley}},\
  and\ \bibinfo {author} {\bibfnamefont {R.~J.}\ \bibnamefont {Van
  Den~Hoogen}},\ }\href@noop {} {\bibfield  {journal} {\bibinfo  {journal}
  {preprint in preperation}\ } (\bibinfo {year} {2022})}\BibitemShut {NoStop}%
\bibitem [{\citenamefont {Hervik}\ \emph {et~al.}(2015)\citenamefont {Hervik},
  \citenamefont {Haarr},\ and\ \citenamefont
  {Yamamoto}}]{hervik2015degenerate}%
  \BibitemOpen
  \bibfield  {author} {\bibinfo {author} {\bibfnamefont {S.}~\bibnamefont
  {Hervik}}, \bibinfo {author} {\bibfnamefont {A.}~\bibnamefont {Haarr}},\ and\
  \bibinfo {author} {\bibfnamefont {K.}~\bibnamefont {Yamamoto}},\ }\href@noop
  {} {\bibfield  {journal} {\bibinfo  {journal} {Journal of Geometry and
  Physics}\ }\textbf {\bibinfo {volume} {98}},\ \bibinfo {pages} {384}
  (\bibinfo {year} {2015})}\BibitemShut {NoStop}%
\end{thebibliography}%

\newpage

\end{document}